\documentclass{aa}
\usepackage{graphicx}
\usepackage{amsmath,amssymb}
\usepackage{epsfig}                            
\usepackage{natbib}
\begin{document}
   \title{Doppler imaging of Speedy Mic using the VLT}

   \subtitle{Fast spot evolution on a young K-dwarf star}

   \author{U. Wolter
          \inst{1} 
          \and
          J.H.M.M. Schmitt \inst{2}
          \and
          F. van Wyk \inst{3}
          }

   \offprints{U. Wolter, \\ \email{uwolter@hs.uni-hamburg.de}}

   \institute{Hamburger Sternwarte, Gojenbergsweg 112, D-21029 Hamburg, Germany \\
              \email{uwolter@hs.uni-hamburg.de}
          \and
             Hamburger Sternwarte, Gojenbergsweg 112, D-21029 Hamburg, Germany \\
              \email{jschmitt@hs.uni-hamburg.de}
          \and
             South African Astronomical Observatory, PO Box 9, 
             Observatory 7935, South Africa \\
             \email{fvw@saao.ac.za}
             }

   \date{Received 25 October 2004 / Accepted 16 December 2004}

   \abstract{
We study the short-term evolution of starspots on the ultrafast-rotating
star HD197890 (``Speedy Mic'' \mbox{= BO~Mic}, K~0-2V, 
\mbox{$P_{\mathrm{rot}}=0.380$~d}) based on two Doppler images taken
about 13~stellar rotations apart.
Each image is based on spectra densely sampling a single stellar rotation.
The images were reconstructed by our Doppler imaging code CLDI (Clean-like Doppler imaging)
from line profiles extracted by spectrum deconvolution.
Our Doppler images constructed from two independent wavelength ranges
agree well on scales down to 10\degr\ on the stellar surface.
In conjunction with nearly parallel V-band photometry our observations
reveal a significant evolution of the spot pattern during as little as two stellar 
rotations. 
We suggest that such a fast spot evolution demands care when constructing 
Doppler images of highly active stars based on spectral time series extending
over several stellar rotations.
The fast intrinsic spot evolution on BO~Mic impedes the determination of a surface differential
rotation;
in agreement with earlier results by other authors we determine an upper limit
of  \mbox{$\vert \alpha \vert <  0.004 \pm 0.002$}.

   }

   \maketitle
%
\section{Introduction}
Doppler imaging allows the reconstruction of surface maps of a sufficiently fast 
rotating star. 
To this end Doppler imaging makes use of the deformations passing through the
spectral line profiles as the star rotates.
The resolution of these maps depends on the spectral resolution, 
phase sampling and noise level of the spectra used as input.
By making use of information modulated into the line profiles,
Doppler imaging overcomes the diffraction 
limitations of direct and interferometric imaging techniques.


Time series of Doppler images of the same object yield information on the evolution 
of surface features.
This evolution is usually split up into ``intrinsic'' spot evolution (spot appearance and decay) 
and large-scale spot motion (tentatively related to meridional flows and differential rotation).
Due to the considerable observational effort required for constructing a 
high resolution Doppler image, in most cases only two images are available
of the same object for direct comparison.
As a result,
the different spot
evolution processes can usually not easily be disentangled.

As to intrinsic spot evolution,
observational results on spot lifetimes are so far of a rather singular nature,
concentrating on a few well-observed objects \citep{Hussain02};
short-term studies, covering a few stellar rotations are rare. 
As an example, 
for the intensively studied apparently single star AB~Dor 
(K0V, \mbox{$P_{\mathrm{rot}}=0.51$~d}) only the polar spot appears to survive time spans
of years \citep[e.g.][]{Donati99}.
We estimate other spots or spot groups, extending several dozen
degrees on the surface, to persist longer than about 5~days 
\citep[][as deduced from comparing their Doppler images of AB~Dor]{Donati97a,Donati99}.
Spots of approximately 10\degr\ size (close to the resolution of the available
Doppler images) seemingly emerge and decay on similar timescales,
i.e. over a few days.
However, as e.g. \citet{Donati99} emphasize, the phase coverage of the
compared Doppler images must be checked carefully in order to derive
significant lifetime estimates of individual features.
For the case of the rapidly rotating G~dwarf star
He~699 (\mbox{$P_{\mathrm{rot}}=0.49$~d}) the available Doppler images
\citep{Barnes98} show that the large-scale distribution of
spots or spot groups remains stable over a month, while features of approximately 10\degr\ size
apparently evolve beyond recognition.


Concerning large-scale surface flows,
the presence of slow meridional flows in the solar convection zone (SCZ) is well
established. 
According to helioseismic measurements, these flows
extend at least several Mm deep into the SCZ \citep{Gizon03}.
However,
long-lived sunspots do not show a regular meridional motion;
their meridional displacements are apparently dominated by intrinsic evolution
\citep{Wohl02}.
While there are possible indications of meridional movements of the spot pattern for
a few evolved stars \citep[e.g.][]{Weber01}, at present no compelling observations of
meridional flows exist for solar-like stars other than the Sun.

As to differential rotation, the observational situation is more
encouraging.
However, determining the rotation 
law of sunspots (i.e. their differential rotation) 
requires extended time series of observations.
Due to the intrinsic evolution of sunspots it is impossible to derive their rotation law 
from two solar images alone.
On the other hand, for suitable stars other than the Sun this appears to be possible
due to the
much larger number of spot features available in each surface image.
The most prominent example is again AB~Dor for which \citet{Donati97a} measured its differential
rotation strength, later confirmed by additional observations \citep{Donati99}.

The angular velocity $\Omega$ of sunspot rotation as a function of heliographic latitude $B$
can be approximated by \citep{Beck99}
\begin{equation}
\label{eq:sunspot_rotation_law}
  \Omega_{\mathsf{sunspots}}(B) \approx
     2.9 \cdot \left( 1 - 0.19 \,\sin^2{B}  \right) \ \ \mathrm{\mu rad\,s^{-1}}   \quad .
\end{equation}
The precise coefficients of this rotation law depend on the size and lifetime of the sunspots
considered, however, deviations are at the level of a few percent. With the same accuracy, this law
also describes the rotation of the solar near-surface plasma.

Presently, a rotation law like Eq.~\ref{eq:sunspot_rotation_law} is also tentatively assumed for 
interpreting observations of stars other than the Sun \citep[e.g.][]{Donati97a, Reiners03}.
However,
parameters to describe the ``strength'' of differential rotation can be defined independently 
of the underlying rotation law.
In the following we use
\begin{equation}
\label{eq:diffrot-alpha_definition}
  \alpha =   \frac{\Omega_{\mathsf{eq}} - \Omega_{\mathsf{pole}}} {\Omega_{\mathsf{eq}}} \quad .
\end{equation}
For the Sun, this definition results in a value of \mbox{$\alpha_{\sun} \approx 0.25$}.
Whether $\alpha$ is a significant physical parameter to describe the dynamics of the outer convection zone
is unclear at present. 

\section{The target ``Speedy Mic''}
Attention was drawn to \object{BO~Mic} by \citet{Bromage92}, reporting a large flare 
observed during the EUV all-sky-survey of \textsc{Rosat};
in retrospect it turned out to be the largest stellar flare
observed during the whole mission.
Their optical follow-up observations yielded
a radial velocity of $-6.5\pm2.0$~km/s
with ``no evidence for binarity'' 
and an estimated $v\,\sin{i}$ of $120\pm20$~km/s,
earning BO~Mic its nickname ``Speedy Mic''.
A further analysis of the same photometric data by \citet{Anders93} did not result in a
well-defined rotation period; this issue was settled by \citet{Cutispoto97},
who determined a value of $0.380\pm0.004$~days.

In summary, its short rotation period makes BO~Mic a promising target for 
Doppler imaging studies because a full stellar rotation can be observed during
a single observing night, allowing the reconstruction of the complete stellar 
surface. 
In this way the influence of intrinsic spot evolution \textit{during} the observations
required for a complete Doppler image can be minimized.

A set of Doppler images, based on observations in July 1998 has
been published by \citet{Barnes01}. As the authors state,
the rather inhomogeneous SNR and phase coverage of their
spectra did not allow detailed statements about the small-scale
spot evolution or differential rotation on BO~Mic.

\subsection{Stellar parameters}
\subsubsection{Magnitudes and colours}
The unusually fast rotation of BO~Mic for an apparently single dwarf star
suggests a young evolutionary status.
Actually, the photometric colours observed by Cutispoto are not consistent with a 
main sequence star \citep{Cox00}.
The observations of \citet{Cutispoto97} yielded a visual magnitude 
at maximum brightness of
\begin{equation}
  \overline{m_V}\approx9.32\pm0.005     \quad.
\end{equation}
Together with the \textsc{Hipparcos} distance of 44.5$\pm$3.2~pc, 
implying a distance modulus of \mbox{3.24$\pm$0.16~mag},
this results in an absolute visual magnitude of
\begin{equation}
  \overline{M_V}\approx6.1\pm0.1    \quad .
\end{equation}
The colours of BO~Mic determined by \citet{Cutispoto97} at brightness maximum
are \mbox{$B-V = 0.92\pm0.005$} and \mbox{$V-R = 0.56\pm0.005$}.
 
\subsubsection{Evolutionary status}
Using the evolutionary models of \citet{Siess00}, including their colour-calibrations,
the above values of $M_V$, $B-V$ and $V-R$ can be put into a consistent
picture of the evolutionary status of BO~Mic. 
Using models of 0.9$\pm$0.05 solar masses we find

\begin{tabular}{ll}                     
    \noalign{\smallskip}
       \hspace*{1cm}   Age        &  3.3$\pm$0.5 \ $\cdot$10$^7$~yr  \\*
       \hspace*{1cm}   Radius     &  0.9$\pm$0.05~$R_{\sun}$      \\*
       \hspace*{1cm}   $T_{\mathrm{eff}}$ & 4750$\pm$50~K           \\*
    \noalign{\smallskip}
\end{tabular}

\noindent 
With  $T_{\mathrm{eff}}$ denoting the effective temperature.
The errors are determined from the evolutionary models based on the photometric errors
given above.
They should be considered as rough estimates because of uncertainties of the models and the
fact that the maximum brightness state observed by   
\citet{Cutispoto97} does presumably not represent a completely spotless hemisphere
of BO~Mic.

The above radius estimate of 0.9~$R_{\sun}$  for BO~Mic resulting from the \citet{Siess00}
evolutionary models is roughly consistent with an estimate based on
the observed rotation of BO~Mic.
An inclination angle of $70\pm10\degr$\ 
and a $v\,\sin{i}$ of 134$\pm$10~km/s, 
both discussed in Sec.~\ref{sec:DI_CLDI}, 
result in an equatorial rotation velocity 
of~$v_{eq}$=143$\scriptstyle{+23} \atop \scriptstyle{-17}$~km/s.
This further corresponds to a radius of~$R=$1.07${\scriptstyle{+0.19} \atop \scriptstyle{-0.14}}~R_{\sun}$,
using the rotation period of 
0.380$\pm$0.004~days.
In light of this,
an inclination at the upper limit of the given inclination range \mbox{($i = 60-80\degr$)}    
and a projected rotational velocity at the lower end of \mbox{$124-144$~km/s}    
seem the be the most likely parameters.

\subsubsection{Lithium abbundance and kinematics}
The Li\,$\lambda6708$ equivalent width of~220~$\pm$50~m\AA\ determined by
\citet{Bromage92}
translates into a Li abundance of 2.3$\pm$0.4~dex, assuming $T_{\mathrm{eff}}$=4800~K. 
\citep{Soderblom93c}.
  Because of the large scatter of Li-equivalent widths and abundances 
  for individual stars, this does not allow to closely constrain the age of BO~Mic.
  As an example, the  Li\,$\lambda6708$ equivalent width ranges from 
  about 100 to  1000~m\AA\ at an effective temperature of 
  4800~K for the Plejades cluster, \mbox{age $\approx$ 70\,Myr}
  \citep{Soderblom93c}.
For members of M34 (age $\approx$ 200~Myr) an abundance of 2.3~dex
is approximately the upper limit detected at this effective temperature
while for members of the Hyades cluster (age $\approx$ 600~Myr) no lithium is
significantly detected at effective temperatures below about 5400~K.
In summary, the Li\,$\lambda6708$ equivalent width of BO~Mic indicates an age below
``a few'' 100~Myr which is compatible with the 
above value of 30~Myr, but does not help to constrain it any further.

 $\,$\citet{Montes2001} analyze the space velocity of a large sample of stars to check their
membership of young stellar kinematic groups;
their results concerning the membership of BO~Mic in the Local Association 
(Pleiades moving group) are inconclusive, yielding no further hint of its age.

\section{Observations and data reduction} 
%



\subsection{Spectroscopy}
The spectral observations were performed at the VLT 
at the ESO Paranal using the spectrograph \textsc{Uves} (Ultraviolet and visual
Echelle spectrograph).
\textsc{Uves} was used in a \textit{dichroic mode},
the spectral ranges thus covered were 3260~\AA\ to 4450~\AA \ and 4760~\AA\ to 6840~\AA\
for the blue and red arm, respectively. 
This permitted  the observation of the ranges above 6000~\AA, well suited for Doppler imaging, and e.g. of
the Ca\,\textsc{ii}~H\&K lines, useful for diagnostics of chromospheric activity.
We have not used wavelength ranges below  6000~\AA\ for the Doppler
images presented here: The massively increased density of
spectral lines for a K-star leads to severe line blending due to rotational
broadening for an ultrafast rotator like BO~Mic.  

The spectrograph slit width was chosen to correspond to 1" on the sky, offering a 
spectral
resolution of $\lambda / \Delta\lambda\approx40\,000$,
adequate for the intended Doppler imaging. 

BO~Mic was observed continuously from dusk to dawn during the two VLT nights
(2002 August~2: JD~2452488.53 to JD~2452488.93, August~7: JD~2452493.51 to JD~2452493.92)
with the exception of a few reference star exposures.
Only the spectra of \object{Gl~472} were used as a template in conjunction with 
the presented Doppler images.

The exposure times of BO~Mic were adjusted between 140~s and 200~s depending on the 
observing conditions
to avoid overexposing the CCDs.
The raw exposures were summed in pairs to obtain the spectra finally used for
the Doppler imaging. The resulting SNR for both nights ranged
typically between 300 and 400 after rebinning to a binwidth of 0.08~\AA.

The reduction of the spectra was carried out in the \textsc{Iraf} environment.
After pre-processing the CCD images and removing scattered light,
the echelle orders were extracted by optimum extraction \citep{Horne86}.
Optimum extraction is primarily intended for the extraction of ``noisy'' spectra,
this would not be necessary in this case (even for the poorest spectra the
SNR exceeded 200 after rebinning).
However, its adaptive modelling of the order extraction apertures 
could deal efficiently with the spectrum variations induced by the
significantly varying seeing conditions (between 0.8\arcsec and 2.5\arcsec). 

One of \textsc{Uves}' CCDs (the MIT-CCD of the ``red arm'') 
shows strong variations (up to about 15\%) in sensitivity on scales
of a few pixels; these variations are visible on technical
flatfields as a ``brickwall pattern''.
In order to adequately correct for these small-scale sensitivity variations 
it was crucial to individually extract a
flatfield for each spectrum using its own extraction aperture.

\subsection{Photometry}
The photometric observations were carried out by FvW 
using the \mbox{0.5m-telescope} and modular photometer 
at the Sutherland site of SAAO \citep{Kilkenny88}.
The standard SAAO reduction procedures were followed, described in the appendix 
to \citet{Kilkenny98}.

BO~Mic was observed continuously during 6~nights (\mbox{2002 August~3--5}, September~12, 13, 15),
as far as weather conditions allowed. Unfortunately, the August~2 and August~7 nights were lost due
to bad weather at SAAO which inhibited almost simultaneous photometry
with our spectral time series.
Two E-region standards \citep{Menzies89},
HD192844 and HD193132,
were used as comparison star and check star, respectively.
The estimated error of the differential photometry is 0.005~mag.

The photometry did not reveal any flares during the observed time span.
The September~12--15 lightcurve exhibits 
the same maxima and minima within \mbox{$\pm$0.01~mag} as
its August~3--5 counterpart (Fig.~\ref{fig:lcurves_folded-SAAOphot_P0380}).
Due to the long time interval between the August and September photometry,
corresponding to about 95~rotations of BO~Mic,
the latter did not add any new information related to the rotation period or
the presented Doppler images.
As a result, the September photometry of BO~Mic is not further discussed in
the following.

\section{Line profile extraction using sLSD}
\begin{figure}
\center{
 \epsfig{file=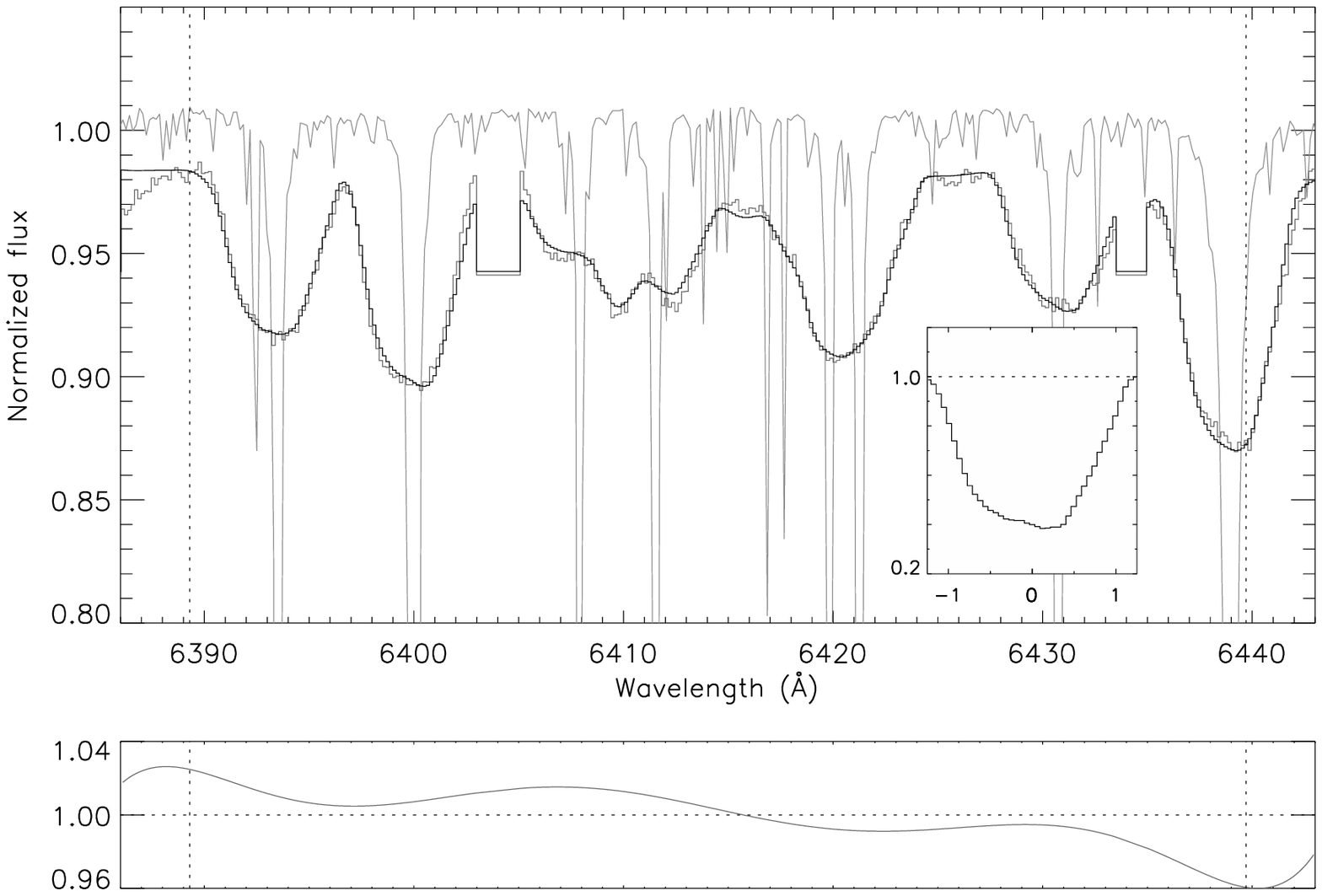, width=1.0\linewidth,clip=} 

\medskip
  \epsfig{file=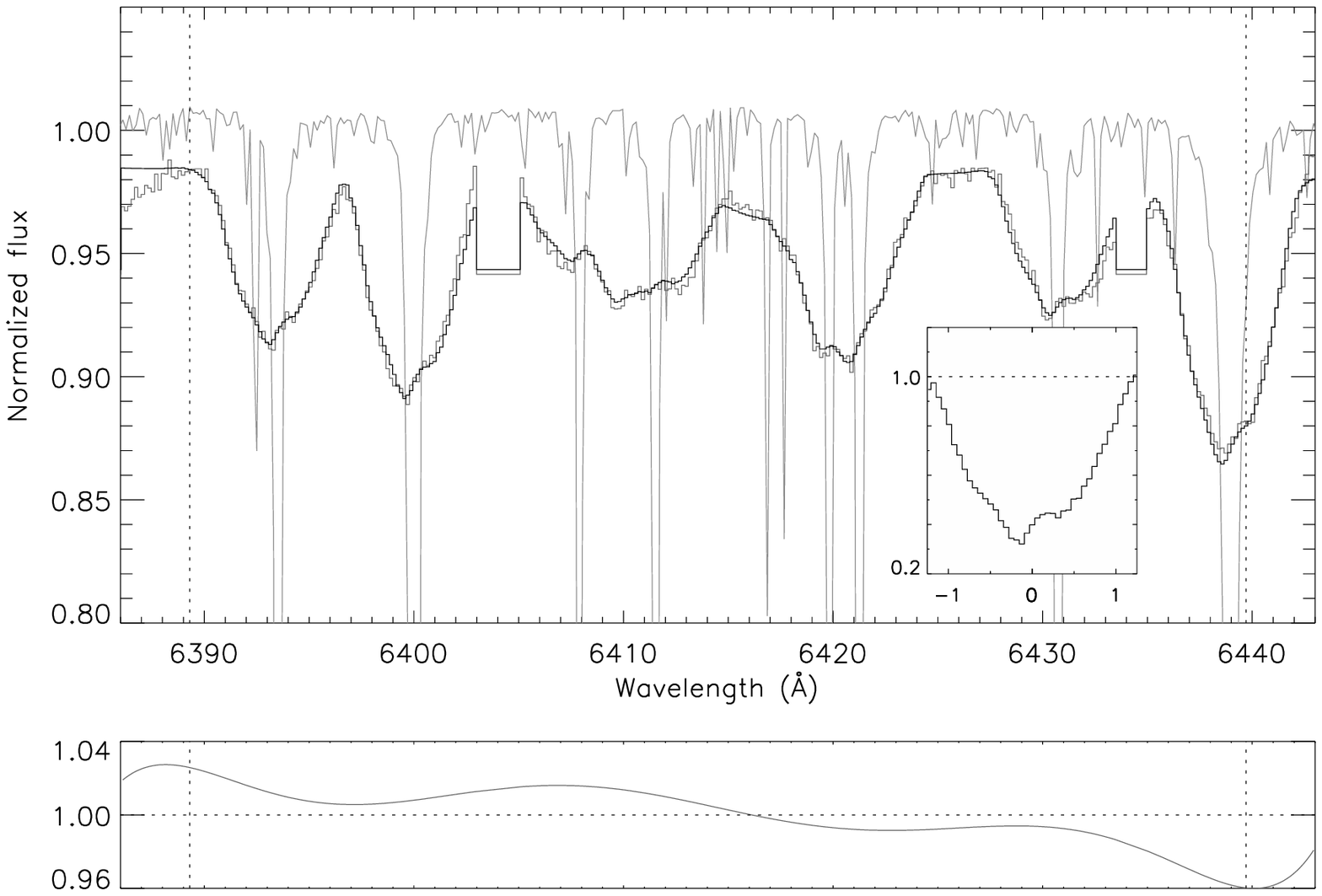, width=1.0\linewidth,clip=}
}
\caption{
Application of spectrum deconvolution (sLSD) to the ``6400\AA'' wavelength region 
of two spectra of BO~Mic.
In the large panels
the gray broad-lined curve represents the observed spectrum, the black spectrum 
shows the fit by sLSD.
The narrow-lined spectrum is the template used for deconvolution
(a synthetic \textsc{Phoenix} 5200\,K spectrum without rotational broadening).
The inset panels show the resulting line profiles (broadening functions), 
their x-axes are annotated in units of 120~km/s.
The lowermost panel shows the ``continuum correction function'' (CCF) used for both 
spectra; see text for details.
        }
\label{fig:sLSD_6400}
\end{figure}
\begin{figure}
\center{
  \epsfig{file=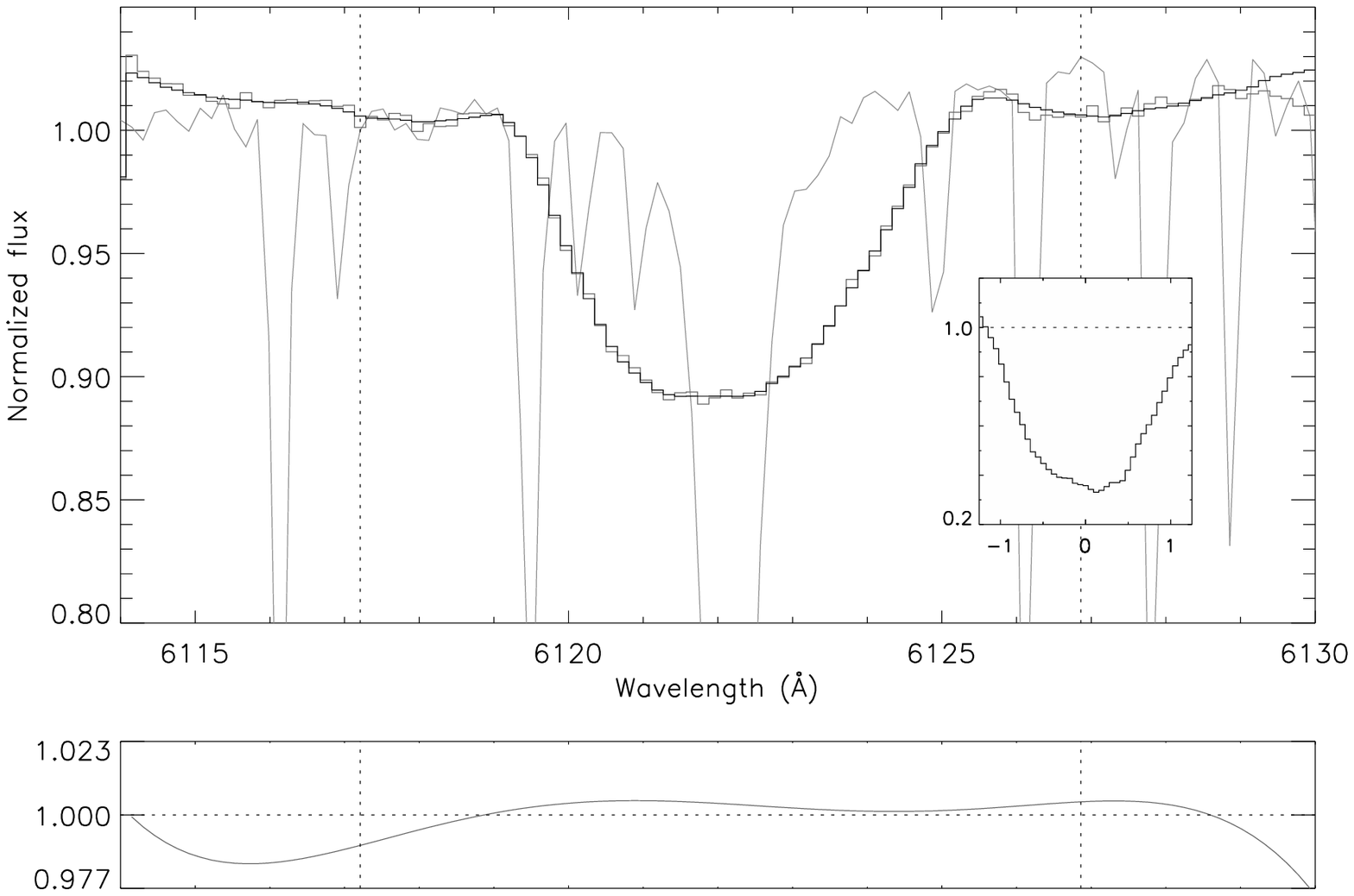, width=1.0\linewidth,clip=}  

\medskip
  \epsfig{file=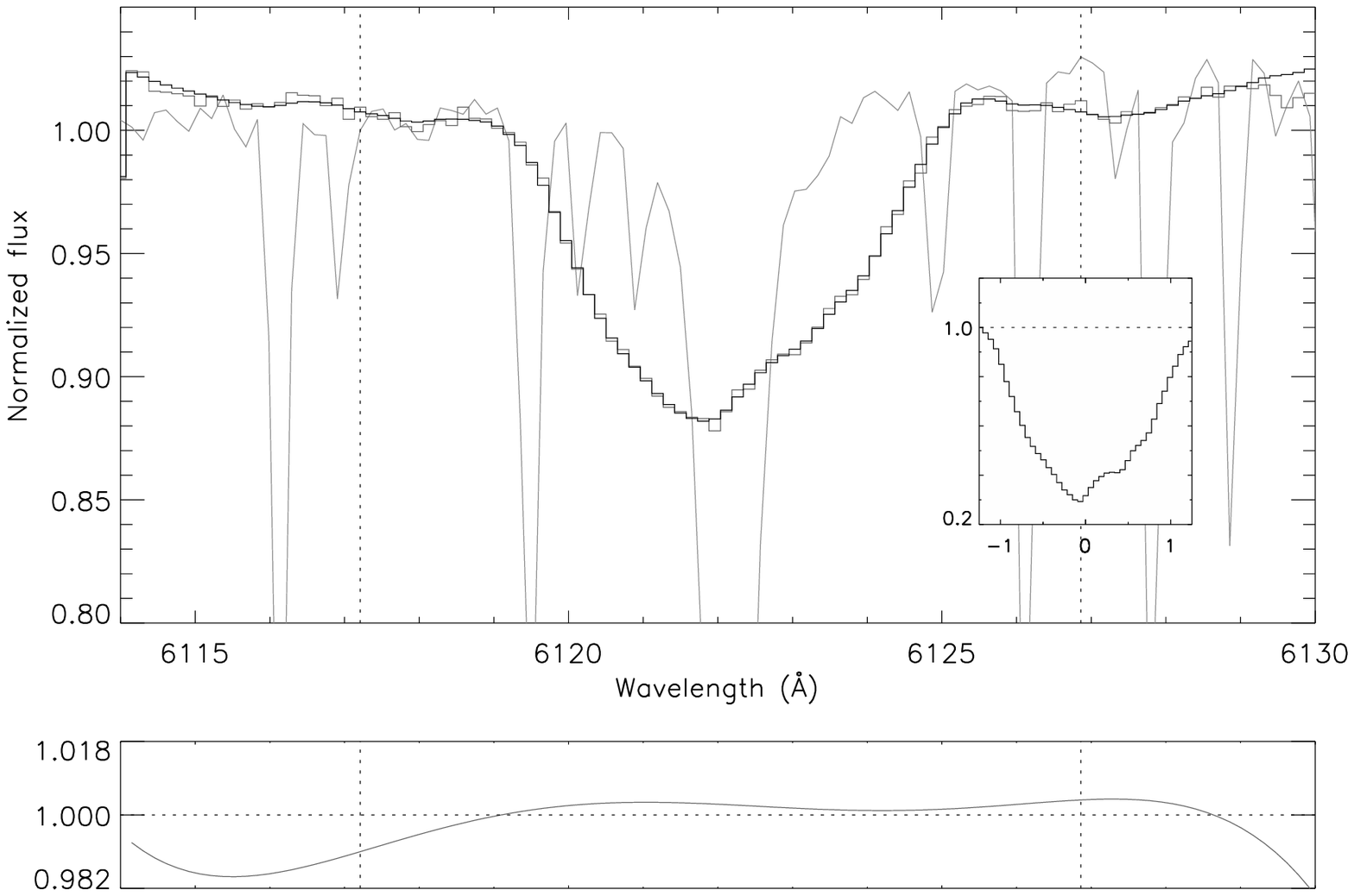, width=1.0\linewidth,clip=}  
}
\caption{ Application of sLSD to the
         ``6120~\AA'' wavelength region; the input spectra are the same as in
         Fig.~\ref{fig:sLSD_6400}, 
         observed on 2002 August\,2 at rotation phases 0.337 and 0.639, respectively. 
         For this wavelength range an observed template spectrum (Gl\,472) was used.
        }
\label{fig:sLSD_6120}
\end{figure}

Deconvolving spectra with the aim of extracting ``average line profiles''
for Doppler imaging was introduced by \citet{Donati97b}.
Originally this was done in the context of magnetic Doppler imaging to 
allow the detection of
the extremely weak signatures of polarization components.
To this end \citet{Donati97b} apply a deconvolution to wide spectral ranges
(e.g. spanning more than 1000~\AA)
with the aim of massively increasing the SNR of the extracted line profiles
compared to individual profiles from the input spectrum.

Donati et al. coined the name ``least-squares deconvolution (LSD)''
for their implementation of this idea.
We call our variant of the same idea 
``\textit{selective} least-squares deconvolution (sLSD)''.
In contrast to the applications of Donati et al., sLSD concentrates
on the analysis of comparatively narrow wavelength ranges, typically
several 10~\AA\ wide, containing only a handful of ``strong'' lines.
While this comes at the price of requiring a higher SNR of the
input spectra, sLSD allows to take the 
local characteristics of the used spectral range into account.

\subsection{The method}
\label{sec:sLSD_method}
The spectral line profiles of a sufficiently fast rotating star can
be approximated by convolving the narrow-lined spectrum of a 
slowly rotating star (called \textit{template spectrum}) with a suitable 
broadening function.
The broadening function describes profile shaping mechanisms not
confined to small regions of the stellar surface:
rotation, limb darkening, macroturbulence and spots.
Using this terminology, sLSD tries to approximate a broadening function
common to the spectral lines of the selected 
wavelength region.

The description of LSD by
\citet{Donati97b}
focusses on a matrix, i.e. a linear formulation of the deconvolution problem.
Presently sLSD does not make use of the linearity of the problem, but
uses a  Levenberg-Marquardt algorithm instead \citep{Press92}.
As in the case of Donati et al.'s LSD, the broadening functions of sLSD are 
discretely sampled and the value of each sampling bin is treated as 
one parameter to be optimized.
It may appear as a disadvantage of sLSD's not-explicitly-linear treatment 
that it does not allow a simple error propagation  
and does not yield a formal error estimate as a byproduct, in contrast
to a linear treatment \citep[e.g.][]{Barnes04}.
However, we think that this disadvantage is not very significant,
because the errors of the broadening functions, which result from the
deconvolution, are clearly dominated by non-statistical errors.
This is supported by the results of \citet{Barnes04} whose
``deconvolved profiles'' (corresponding to sLSD's broadening functions)
show a ``rippling effect'', most pronounced in the continuum regions
surrounding the broadening functions, with an amplitude
massively exceeding  their estimated statistical errors.
Because of this ``rippling effect'' we have actually excluded these
continuum regions around the broadening functions from the deconvolution.  

A truly significant error estimate of the broadening functions
resulting from sLSD (or LSD) presently appears as an open question. 
We think that such an error estimate would require a systematic study
partly based on synthetic input data, e.g. along the lines of \citet{Barnes04}.
However, in addition to \citet{Barnes04}, mismatches of the template spectrum
presumably need to be considered: They increase the ill-posedness of
the deconvolution problem of sLSD (or LSD), leading to instabilities
of the resulting optimization process which could be the reason of
the ``rippling effect'' mentioned above.

A different parametrization of the broadening functions, 
e.g. by Chebichev polynomials, can be easily implemented in sLSD. 
However, in practice this does not reduce the number of parameters 
needed to describe the broadening function: The  proper approximation of the
large curvature at the transition into the surrounding continuum
requires unwieldy high order polynomials.

\subsection{Application and ingredients}
\label{sec:sLSD_app}
For BO~Mic we selected
two wavelength regions for the extraction of line profiles by sLSD.
They were intentionally chosen as  quite different in character
(Figs.~\ref{fig:sLSD_6400} and~\ref{fig:sLSD_6120}): 
While the ``6120\,\AA''-region 
is dominated by a single line (Ca\,\textsc{i}\,$\lambda$6122), the ``6400\,\AA''-region
comprises a handful of stronger lines (the strongest are Fe\,\textsc{i}\,$\lambda$6400
and Ca\,\textsc{i}\,$\lambda$6439).

Figs.~\ref{fig:sLSD_6400} and~\ref{fig:sLSD_6120} 
illustrate the application of sLSD to the two wavelength ranges.
In the large panels the gray broad-lined curve represents the observed spectrum, 
the black spectrum shows the fit achieved by sLSD.
The vertical dotted lines delimit
margin regions contributing to the spectrum fit with reduced weights;
the narrow flat regions around 6404~\AA\ and 6434\AA\ 
in Fig.~\ref{fig:sLSD_6400} have been excluded from  the fit (see
below in Sec.~\ref{sec:sLSD_CCF}).

The narrow-lined spectra show the templates used for the deconvolution.
In case of Fig.~\ref{fig:sLSD_6400} a synthetic spectrum 
generated for an effective temperature of 5200\,K 
by the spectrum synthesis code \textsc{Phoenix} \citep{Hauschildt99} has been used.
The deconvolutions shown in Fig.~\ref{fig:sLSD_6120} make use of an observed
template spectrum (Gl~472, K1V); spectra synthesized by \textsc{Phoenix}
tentatively adopting a wide range of element abundances and effective
temperatures yielded significantly poorer spectral fits in this 
wavelength region.
 

The application to relatively narrow wavelength intervals leads
to several issues to be treated by sLSD:
The treatment of end-effects at the borders of the wavelength
interval, tentatively correcting for small template deficiencies
and stabilizing the convergence of the broadening function optimization.
These issues are discussed below in Secs.~\ref{sec:sLSD_CCF} and~\ref{sec:sLSD_regu}.

\subsubsection{End-effects and template deficiencies}
\label{sec:sLSD_CCF}
\begin{figure}
\center{
  \epsfig{file=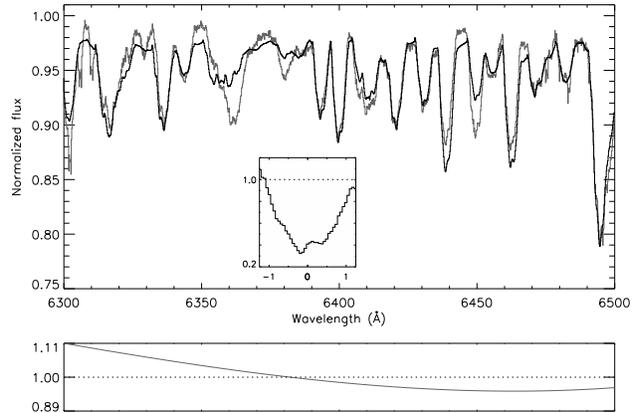, width=1.0\linewidth,clip=}  
}
\caption{
         Application of sLSD to a moderately wide spectral range around the ``6400\AA'' wavelength region.
         The input and template spectra are the same as in the lower panel of Fig.~\ref{fig:sLSD_6400},
         the latter is not shown here.
         The reasons for the rather poor fit to the spectrum are discussed in the text.
        }
\label{fig:sLSD_6300-500widerange}
\end{figure}
%
Fig.~\ref{fig:sLSD_6300-500widerange} shows an example
application of sLSD using the same template as in 
Fig.~\ref{fig:sLSD_6400}, but yielding a poorer fit to the input spectrum.
This rather poor spectral fit 
is due to mismatches with the template spectrum (leading to slightly
wrong equivalent widths of some lines, most pronounced in the region around 6360~\AA) and
regions containing echelle order overlaps not optimally merged (the region near 6460~\AA).
To avoid the resulting deficiencies of the fit shown in Fig.~\ref{fig:sLSD_6400},
three measures have been taken: 
(i)~Narrowing down the wavelength region, carefully selecting the  boundaries of 
the fitted wavelength range and excluding narrow intervals from the spectral fit
(near 6403~\AA\ and 6435~\AA).
(ii)~Defining margin intervals at the boundary of the fitted range 
where the relative weights 
used for the $\chi^2$-optimization of the spectral fit are reduced;
these margin intervals are marked by vertical lines in 
Figs.~\ref{fig:sLSD_6400} and~\ref{fig:sLSD_6120}.
(iii)~Introducing an
adaptive ``continuum correction function'' (CCF), shown in the lower
narrow panels of Figs.~\ref{fig:sLSD_6400} to~\ref{fig:sLSD_6300-500widerange}.

The exclusion of narrow wavelength intervals from the spectral fit, mentioned in
item~(i), is motivated by template spectrum
mismatches, i.e. apparently missing weak lines in these intervals.
These missing lines in the synthetic template spectrum may be due to
wrong element abundances
supplied to \textsc{Phoenix} or wrong atomic data. 
However, we were unable to find element abundances leading to an improved spectral fit,
so wrong atomic data seem more plausible. 
Actually, the exclusion of the intervals is not necessary in this case, 
because it has only a marginal influence on the line profiles resulting from the 
deconvolution. 

Measure~(ii), namely
the margin intervals, would not be necessary if the selected wavelength range
were surrounded by practically line-free regions, i.e. 
intervals of ``undisturbed'' continuum.
However, such undisturbed regions are largely absent in the visible spectra
of very fast rotating cool stars (which are the primary candidates for
Doppler imaging). 
By introducing the margin intervals, sLSD reduces the influence of strong lines 
completely or partly outside the selected wavelength range.

The most important measure to reduce this influence is the application
of the above named CCF.
The observed spectrum is divided by the CCF prior to the optimization
of the broadening function.
The CCF is chosen as a slowly varying function of wavelength,
i.e. varying on larger scales than the individual line profiles of the spectrum.
In this way
the CCF adapts the continua of the observed and template spectra.
This makes a continuum normalization of the input spectra prior to sLSD unnecessary.
Such a normalization is usually very difficult for strongly rotationally broadened spectra.
It remains so when using the CCF, but the process can be monitored during the application
of sLSD, instead of being potentially ``hidden'' in the original spectra reduction.
Additionally the CCF compensates for the effects of strong lines at the boundary
of the wavelength region.
This compensation is clearly visible in Figs.~\ref{fig:sLSD_6400} and~\ref{fig:sLSD_6120}
showing the deformations of the CCF at the boundaries. 
Finally the CCF tentatively corrects for some deficiencies
of the template spectrum 
(wrong equivalent widths or missing weak lines).
An example for such a correction can be seen when comparing the region around
6410~\AA\ in Figs.~\ref{fig:sLSD_6400} and~\ref{fig:sLSD_6300-500widerange}.

The CCF is iteratively constructed by alternating optimization steps
for the broadening function and the CCF.
The CCF is chosen as a Legendre polynomial of fixed degree;
this degree needs to be chosen sufficiently high to provide 
``flexibility'' for the intended corrections.
On the other hand, the degree should be adequately
low to avoid the CCF adapting to deformations of individual line profiles
which are to be fitted by the broadening function.
As an example, the degrees are 7 in Fig.~\ref{fig:sLSD_6400}, 5 in Fig.~\ref{fig:sLSD_6120}
and 3 in Fig.~\ref{fig:sLSD_6300-500widerange},
respectively.

For the extraction of a time series of line profiles the CCF should be kept fixed,
or its constancy closely monitored.
Otherwise variations of the CCF, e.g. due to instabilities of the optimization could 
introduce ``fake'' variations of the extracted line profiles.

\subsubsection{Stabilizing the convergence}
\label{sec:sLSD_regu}
During the early phase of the broadening function optimization
small-scale oscillatory solutions (typically with large bin-to-bin amplitudes)
tend to appear.
This tendency is further enhanced by the alternating optimization
of the CCF and the broadening function.
These oscillatory solutions massively disturb or even prevent the convergence
of sLSD. 
To avoid this,
sLSD employs a Tikhonov-regularization of the optimization of the broadening function;
i.e. it applies a ``penalty function'' to suppress strong small-scale
gradients of the solution.
The smoothing effect of the Tikhonov-regularization on the solution profile 
is adjusted by the weight of the penalty function during the optimization.
This weight is fixed prior to the sLSD iteration; for input spectra of low or moderate 
noise it can be chosen very small. In this way it suppresses the oscillatory solutions
without noticeably smoothing the final solution profiles, i.e. without a
loss of potential starspot signatures in the profiles.

\subsection{Selecting the template spectrum}
\begin{figure}
\center{
  \epsfig{file=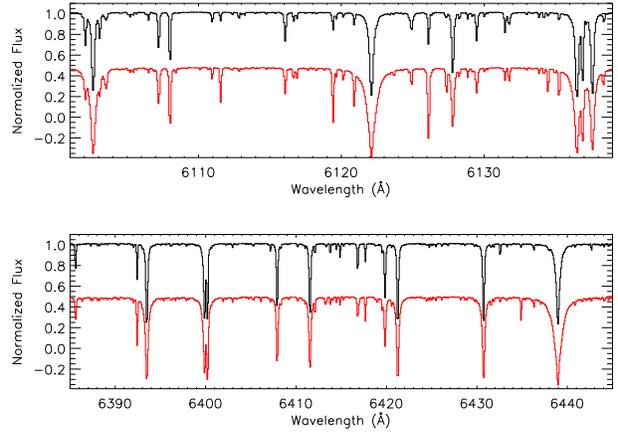, width=1.0\linewidth,clip=} 
}
\caption{
  Selected ranges from synthetic spectra, not rotationally broadened.
  The \textsc{Phoenix} 5200\,K spectrum (black) has been used as the template spectrum for applying
  sLSD to the ``6400~\AA'' wavelength range. 
  The  \textsc{Phoenix} 4600\,K spectrum (red/gray) is considered as a proxy for the spectrum of the spots
  on BO~Mic, see text for details.
  For both panels, the quasi-continua have been normalized to unity, 
  the red/gray spectrum is shifted in flux by -0.5.
        }
\label{fig:phoenix_5200Kvs4600K}
\end{figure}
The observed stellar spectrum results from a disk integration 
and is in effect a weighted sum of
contributions from undisturbed photospheric regions and spots.
As a consequence, (at least) two intrinsic spectra take part in the
disk integration, but only one template is used for the deconvolution.
Since the disk integration is dominated by the spectrum of the undisturbed photosphere,
due to its greater brightness, 
the template spectrum for the deconvolution needs to be selected similar to it.
In this way the intrinsic spectrum (or even different spectra) of the spots
is not included as input information to the spectrum deconvolution.

If the intrinsic spectra of the unspotted and spotted photosphere \textit{were identical},
extracting a broadening function common to the whole stellar surface would be strictly correct.
Naturally, because of the strong temperature contrast and other differing 
atmospheric parameters, these spectra are not identical.
However, Doppler imaging based on spectrum deconvolution apparently 
is quite successful as demonstrated by the works of Donati et al. and
also supported by the results of this paper.
As discussed in the following, 
such a success can only be expected
if the spot-photosphere temperature contrasts are either small or large,
in a sense to be defined below.

Fig.~\ref{fig:phoenix_5200Kvs4600K} illustrates
a spot-photosphere temperature contrast which is small compared
to e.g. the Sun.
The spectrum of lower effective temperature is used here as a tentative proxy
for the spectrum of the spots on BO~Mic,
the temperature contrast has been chosen as 600\,K in this example.
This approximately represents the situation assumed for the construction
of the Doppler images discussed in Sec.~\ref{sec:SM_DIs},
because in the considered wavelength range
the ratio of the quasi-continuum fluxes of the two spectra
is about 0.5.

In this case the dominating lines
in the shown wavelength ranges are very similar for the spots and the undisturbed photosphere.
As a result, the line profile deformation by cold spots
is predominantly due to the spot-induced change of continuum flux
and approximately correctly recovered by the deconvolution with a single template spectrum.
Obviously, this does not apply  to spectral lines strongly varying with temperature;
consequently they should be avoided when using sLSD for the line profile extraction.

If on the other hand a large spot-photosphere temperature contrast is assumed
(e.g. 1800K, as deduced for the highly active K-dwarf LQ~Hya by \citealt{Saar01}),
the flux emitted by the spots is very small compared to the surrounding photosphere
(less than 10\% in the example of LQ~Hya, estimating from 
\textsc{Phoenix} synthetic spectra of corresponding effective temperatures).
In this case 
the intrinsic spectrum of the spots has very little influence on the resulting
broadening function.
Again, the deconvolution based on a single template spectrum should produce 
approximately correct results.

\section{Doppler imaging using CLDI }
\label{sec:DI_CLDI}

\subsection{The method}
\label{sec:CLDI_general}
We call
the Doppler imaging algorithm used in the following
``\textsc{Clean}like Doppler imaging'' (CLDI),
it is based on the algorithm developed by \citet{Kuerster93}.
CLDI is described in detail in \citet{Wolter04}, further descriptions and 
studies of its behaviour
will appear in a separate publication (Wolter, 2005, in preparation). 
CLDI's rationale
is very similar to the algorithm described and tested in \citet{Kuerster93}.

The mathematical formulation of the Doppler imaging problem results in
a matrix equation, \mbox{$D  =  \mathbf{R} \cdot I$} (e.g. \citealt{Vogt87}).
For this formulation, the line profiles are discretely sampled 
and assembled
end-to-end giving a one-dimensional vector $D$. Likewise, the surface has
been divided into a discrete grid of zones 
whose values  
have also been packed into a one-dimensional vector $I$ .
The matrix $\mathbf{R}$ is the so-called \textit{response matrix}. At the given
resolution it describes the observable line profiles as a function of the stellar
surface.

Doppler imaging is an inverse problem making this equation
ill-posed and ill-conditioned (e.g. \citealt{Lucy94}).
In a way, CLDI approaches the 
equation in a mathematically ``daring'' way:
Since the inverse $\mathbf{R}^{-1}$ does not exist,
it attempts a solution of the above equation by iteratively applying the transpose
of the response matrix $\mathbf{R}^{\!\mathsf{T}}$ to the observed 
line profile deformations.
More precisely, it applies a slightly modified
$\mathbf{R}^{\!\mathsf{T}}$ to the difference of the observed 
line profile deformations
and their reconstructed counterparts at each iteration step.
As discussed in \citet{Wolter04}, this iterative ``backprojection'' of the
line profile deformations onto the reconstruction surface 
can be justified by a geometric interpretation of $\mathbf{R}^{\!\mathsf{T}}$. 

CLDI is intended as an alternative to regularized
optimization schemes for the inversion of Doppler imaging problems, 
like maximum-entropy Doppler imaging.
In contrast to methods based on regularization, CLDI does not
explicitly treat Doppler imaging as a continuous optimization
problem. 
Instead it attempts a discrete approximation of the observed line
profile deformations in the way described above.
In this way,
inspired by the appearance of sunspots, CLDI models surface reconstructions 
of high contrast and discrete contrast steps.

\subsection{sLSD and CLDI}
\label{sec:sLSD+CLDI}
\begin{figure}  
\center{
  \epsfig{file=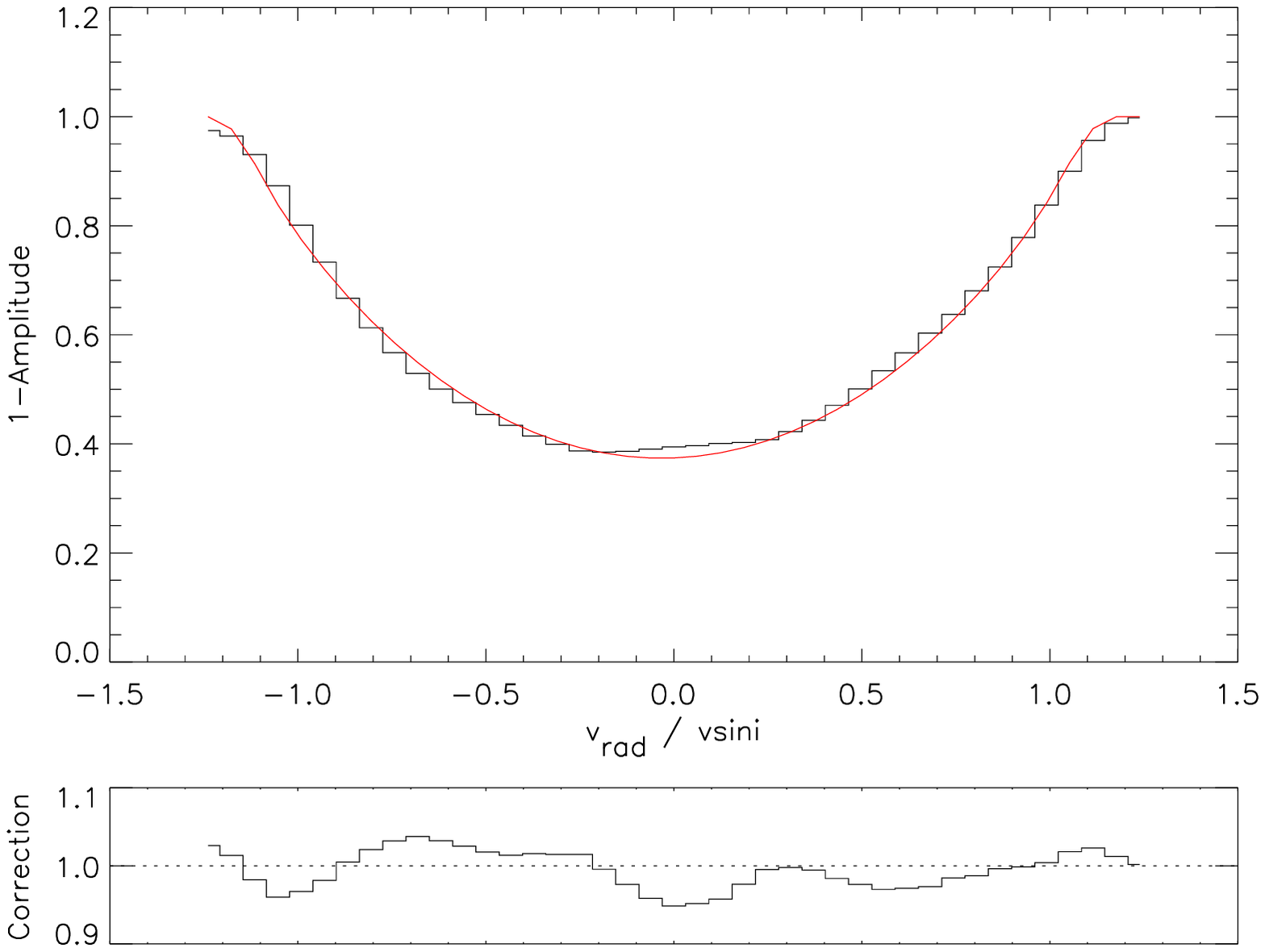, width=0.80\linewidth,clip=} 
  \vspace{-0.4cm}
  \epsfig{file=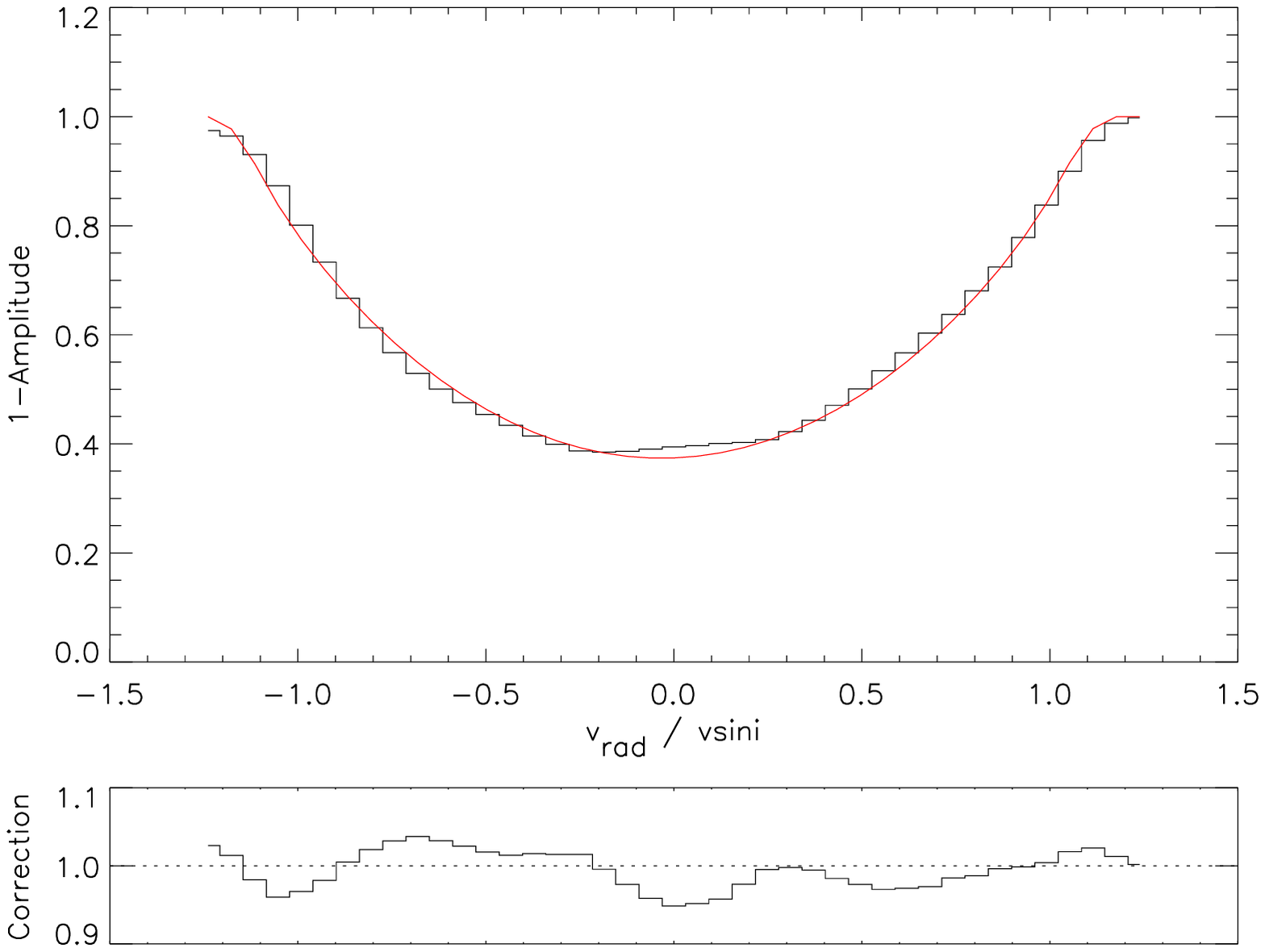, width=0.80\linewidth,clip=} 
}

\bigskip
\caption{
  \textit{Upper panel}:
  Average line profile of BO~Mic (stepped graph) 
  extracted by sLSD from the 
  ``6400\,\AA'' 
  wavelength region;
  the radial velocity is annotated in units of 120~km/s. 
  The smooth graph shows a fitted analytical rotation profile,
  its parameters are given in Table~\ref{tab:CLDI_SM_mainparams}.
  \textit{Lower~panel}: 
  Correction function $\xi$ computed from Eq.~\ref{eq:RBF_correction_xi}.
  See text for details.
        }
\label{fig:sLSD_RBFcorrection}
\end{figure}

The line profiles (more precisely the broadening functions, 
cf. Sec.~\ref{sec:sLSD_method}) extracted 
from average spectra of BO~Mic using sLSD show minor deviations from symmetry,
an example is shown in Fig.~\ref{fig:sLSD_RBFcorrection}.

Phase-\textit{independent} line profile asymmetries
massively disturb the convergence of CLDI. 
The reason is that even weak deviations of this kind mislead CLDI's
``backprojection'' of the input line profiles 
used for approximately locating (groups of) spots on the surface. 
Such deviations are intrinsically impossible to fit for Doppler imaging, because
they lie outside the line center without migrating through the profile with rotation phase.
We therefore have to make sure that the time series of line profiles
is as far as possible free of such phase-independent asymmetries.

Such asymmetries demand care when interpreting the shape of the line profile
in terms of stellar parameters \citep[e.g.][]{Reiners03}.
For the purpose of Doppler imaging these asymmetries are
less crucial, if they are small compared to 
the spot-induced line profile deformations.

In order to ensure the convergence of the Doppler imaging,
the line profiles extracted from the spectra of BO~Mic have been transformed 
to symmetric rotation profiles in the following way,
illustrated in Fig.~\ref{fig:sLSD_RBFcorrection}.
  The shown profile was extracted from an average of all spectra of BO~Mic
  taken on August\,2,
  these spectra sample a whole stellar rotation quite evenly.
  In this way spot-induced line profile deformations should largely cancel out,
  assuming that the spot pattern does not change during the observations.   

A symmetric rotation profile is determined by fitting an 
analytic rotation function $f_\mathrm{rot}$
to an average line profile of the time series.
$f_\mathrm{rot}$ is computed from a ``rotation profile''  $G$ \citep{Gray92}
\begin{equation} 
\label{eq:rotation_profile}
  G(v_\mathrm{rad}) = 
   \frac{   2 (1-\epsilon) \mathsf{A}^{\frac{1}{2}}
         +  \frac{1}{2} \pi \epsilon  \mathsf{A}
        }
        {\pi W (1-\epsilon/3)}    
\end{equation}
by convolving it with a Gaussian of constant width.
Here the symbol $\mathsf{A} \equiv  1 - \left( v_\mathrm{rad} / (c \cdot W)\right)^2$ has been used,
computed from  
\begin{equation*}
  W=\frac{v\,\sin{i}}{c} \cdot \lambda_0    
\end{equation*}
using the center wavelength of the considered line
$\lambda_0$.

A correction function $\xi$ is computed from the average 
line profile $b_\mathrm{ave}$ and the 
corresponding analytic fit $f_\mathrm{rot}$:
\begin{equation}
  \label{eq:RBF_correction_xi}
  \xi(v_\mathrm{rad}) = \frac{f_\mathrm{rot}(v_{rad}) } {b_\mathrm{ave}(v_\mathrm{rad})}
\end{equation} 
A sample correction funtion is shown in the lower panel of 
Fig.~\ref{fig:sLSD_RBFcorrection}, in that case the amplitude of $\xi$ is about 0.03,
i.e. the correction amounts to at most 3\% of the continum surrounding the line profile.
Each individual line profile of the time series is then corrected by applying \textit{the same}
correction function
\begin{equation}
  \label{eq:RBF_correction_applied}
 b_\mathrm{corr}(v_\mathrm{rad}) = b(v_\mathrm{rad}) \cdot \xi(v_\mathrm{rad})
\end{equation} 
The thus corrected $b_\mathrm{corr}$ are used as the input to CLDI.
Using the same correction $\xi$ for the whole time series is a natural requirement; 
it avoids introducing artificial phase-dependent deformations of the line profiles, which could 
cause artefacts in the resulting Doppler images.

Care must be taken that \textit{symmetric} deviations between the average
input profile $b_\mathrm{ave}$ and $f_\mathrm{rot}$ are not ``corrected away'' 
during this process. Such symmetric deviations are induced
e.g. by \textit{polar spots} (and other rotationally symmetric spot configurations).
As illustrated by Fig.~\ref{fig:sLSD_RBFcorrection},
such signatures were not
detected for the average profiles of BO~Mic.

\subsection{The images}
\label{sec:SM_DIs}
\begin{figure*}
\epsfig{file=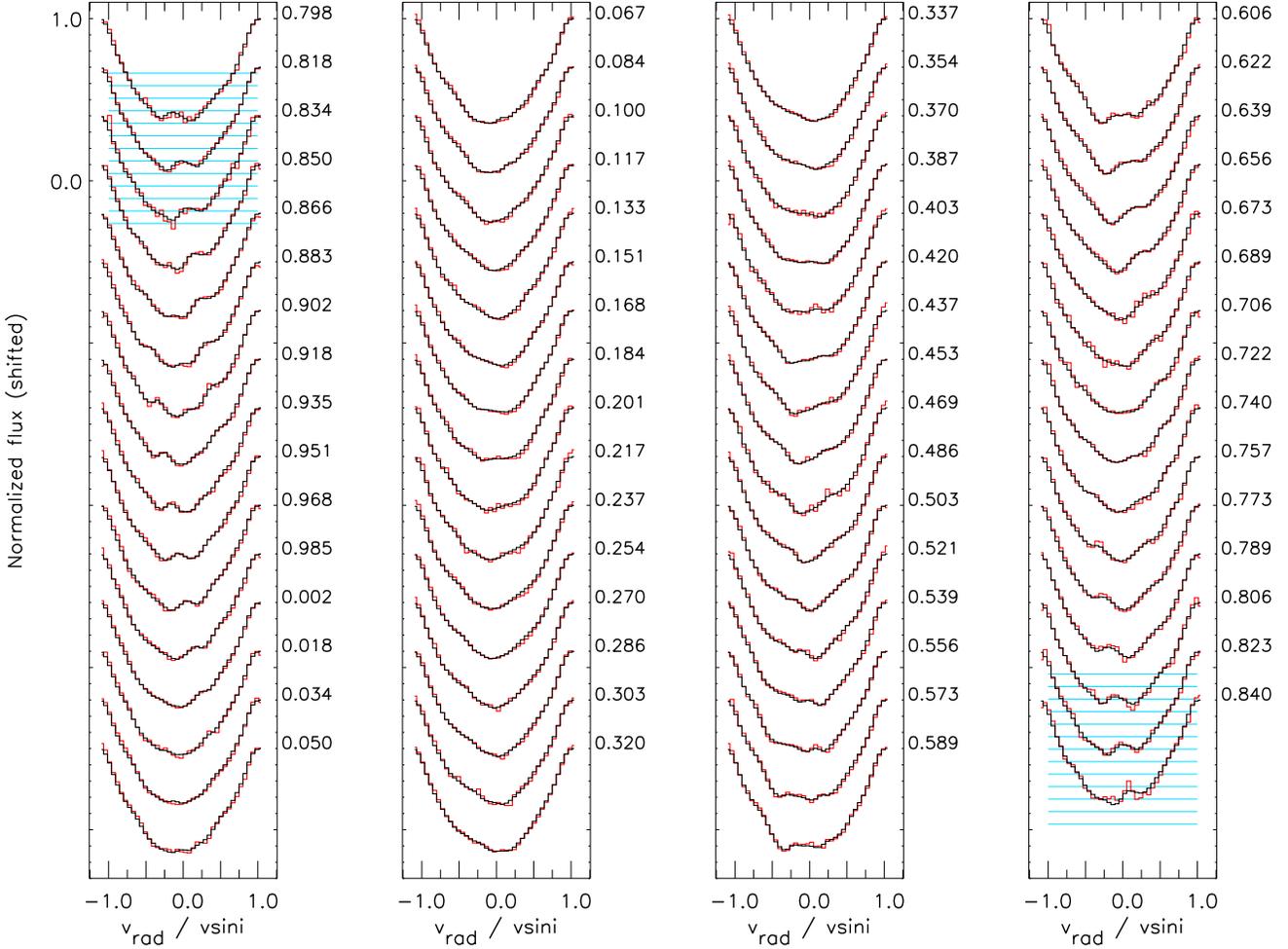,width=0.95\linewidth}
  \caption{Line profiles of BO~Mic, extracted by spectrum deconvolution in the ``6400\AA''-region~(red/gray)
           and fitted by CLDI (black). 
           The reconstruction parameters are listed in Tab.~\ref{tab:CLDI_SM_mainparams}.
           Subsequent profiles of the time series are shifted, 
           the flux scale is valid for the topmost profile of each panel.
           The rotation phases
           are given right of each profile,
           with the phase zero-point at $\mathsf{JD}_{\phi=0}= 2448000.05$ 
           Both observation nights were longer than the rotation period of 0.380~days,
           the hatched regions mark the 
           cores of the profiles in the phase overlap interval
           of the beginning and the end of the night.
           }
  \label{fig:sLSD6400-CLDIvssLSD_1st}
\end{figure*}

Two independent time series of line profiles were extracted from the spectra of 
BO~Mic (Sec.~\ref{sec:sLSD_app}),
the corresponding images are denoted by ``6120\,\AA'' and ``6400\,\AA''
in the following.
The images are shown in Fig.~\ref{fig:SM-CLDI-RMxTSA_HA-P0380-truecoords-1st2nd},
the achieved line profile fits are shown for one of them in
Fig.~\ref{fig:sLSD6400-CLDIvssLSD_1st}.

As stated above, the SNR of the input spectra ranged typically between
300 and 400, for some spectra it reached up to 500. Only for 3 of
the spectra was the SNR below 250. We have carried out tests of CLDI using synthetic input
profiles with very similar reconstruction parameters as those used for
our reconstructions of BO~Mic (Wolter, 2005, in preparation).
These tests suggest a ``critical'' SNR range for the Doppler imaging
reconstructions,
mainly depending on the resolution aimed at and on the adopted
atmospheric parameters.
When the SNR of the input profiles falls below this critical range,
a serious deterioration of image details results.  
Above this level, the reconstruction quality only increases weakly
with increasing SNR.
Our tests indicate that this critical SNR range lies between about
200 and 250 for the parameters adopted here. 

\subsubsection{Image reconstruction parameters}
The image reconstruction parameters are summarized in Table~\ref{tab:CLDI_SM_mainparams}.
The limb-darkening and $v\,\sin{i}$  values 
have been determined by fitting analytical
rotation profiles  $f_\mathrm{rot}$ to average line profiles (broadening functions) 
of BO~Mic (Sec.~\ref{sec:sLSD+CLDI}).
It should be noted, that the errors of the fit parameters
are presumably larger than the margins given in the table.
This is mostly due to the asymmetries of the average line profiles, introducing 
uncertainties into the 
fit.\footnote{
  Analyzing the same profiles in Fourier space \citep{Reiners02a}
  yields \mbox{$v\,\sin{i}\approx131\pm2$~km\,s$^{-1}$}. 
  However, since this method is also affected by problems of profile asymmetry,
  the error may well be larger.
             }

\begin{table}[h]
  \center
  \caption[]{Parameters adopted for the construction of the Doppler images
             of BO~Mic,
             $\epsilon$ represents the linear limb darkening coefficient.
             The spot continuum flux is given relative to an undisturbed 
             photospheric continuum flux of one.
             Note that the given margins designate the deviations between parameters 
             adopted for different
             reconstructions, they are not proper error estimates.
            }
  \label{tab:CLDI_SM_mainparams}

  \smallskip
  \begin{tabular}{ll}                     
    \hline
    \noalign{\smallskip}
    Rotation period          &  $P_0$=0.380\,days  \\
    Inclination              &  70$\degr$ \\
    Spot continuum flux      &  0.5\\
    $v\,\sin{i}$              &  134$\pm$2~km\,s$^{-1}$  \\
    $\epsilon$                       &  0.9$\pm$0.1 (``6120~\AA'')  \\
                              &  0.7$\pm$0.1 (``6400~\AA'')  \\
    \noalign{\smallskip}
    \hline
  \end{tabular}
\end{table}

The adopted value of 0.5 for the spot continuum flux means that a surface element completely 
covered with spots (i.e. with a spot filling factor of one) emits half of the flux in the 
continuum around the synthesized line profile, compared to an uncovered surface element
(spot filling factor zero).
Adopting a lower value, i.e. a larger spot-photosphere contrast, leads to 
a slightly more ``jagged'' appearance of the spot groups of the
resulting Doppler images,  
it has no significant influence on the reconstructed spot pattern.

\subsubsection{Inclination}
\begin{figure}[h] 
  \epsfig{file=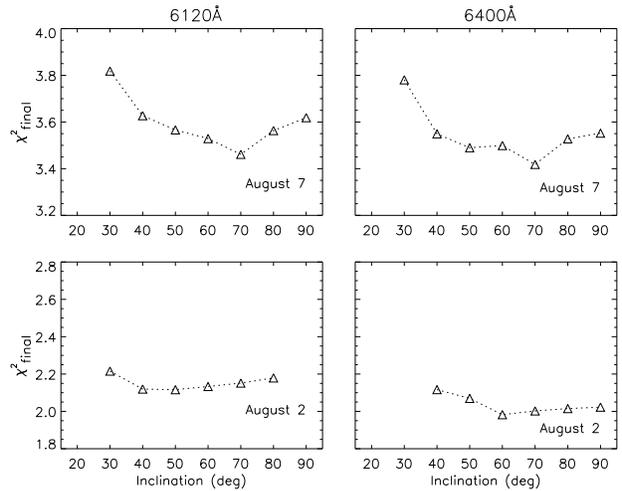,width=\linewidth,clip=}     
  \caption{
           Quality of line profile fits $\chi^2_\mathsf{final}$ achieved by CLDI
           for different datasets of BO~Mic
           as a function of the adopted inclination of the reconstruction star.
           The wavelength ranges and
           observation dates corresponding to each graph are annotated.
          }  \label{fig:SM_CLDI_inc-chisqfin_P0380} 
\end{figure}

Adopting an inclination for Doppler imaging is commonly done by attempting
reconstructions at different inclinations and selecting the one
yielding the best line profile fits.
Fig.~\ref{fig:SM_CLDI_inc-chisqfin_P0380}
shows the line profile fit quality expressed as $\chi^2_\mathsf{final}$
as a function of the adopted inclination for different datasets of BO~Mic.
These curves do not show a strong variation of $\chi^2_\mathsf{final}$;
this behaviour is not typical of CLDI, as we have tested for surfaces 
reconstructed from synthetic data.

The minima of $\chi^2_\mathsf{final}$ 
for the August~7 reconstructions coincide at $70\degr$.
The August~2 reconstructions show less pronounced minima at lower values
which do no agree.
As discussed above, the radius of BO~Mic tentatively deduced from evolutionary 
models clearly suggest an inclination $\ga 70\degr$´.
As a result this value has been chosen for the reconstructions shown here.
Actually varying the inclination by as much as $\pm20\degr$ has only little
influence on the resulting Doppler images.

\begin{figure*}  
\begin{tabular}{c|c}
 \parbox{0.45\linewidth}{
  \center{\textsf{August 2 - ``6120~\AA''}} \\ 
  \vspace*{-0.5cm}
  \includegraphics*[width=\linewidth,clip=]{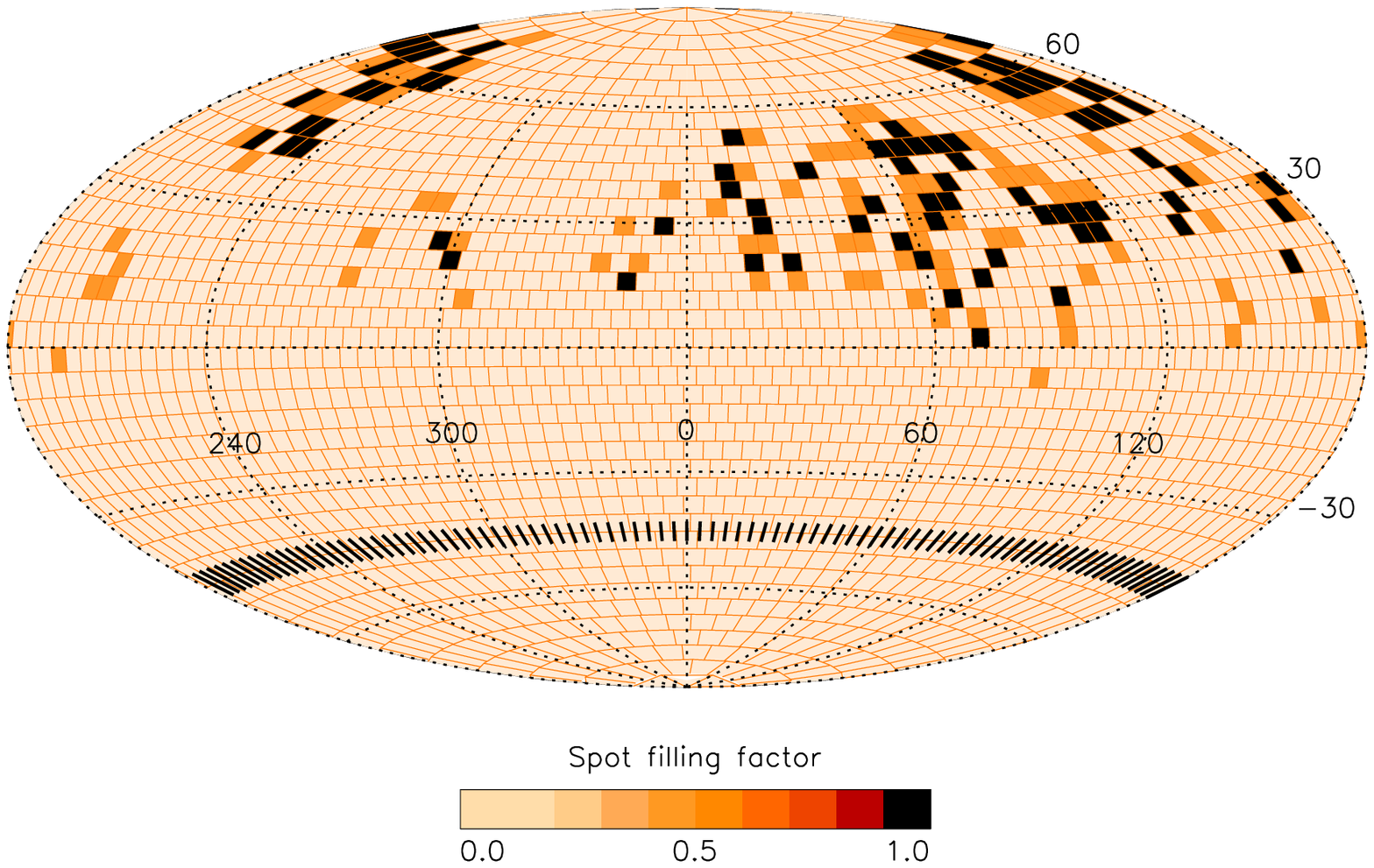}   
  \includegraphics*[width=\linewidth,clip=]{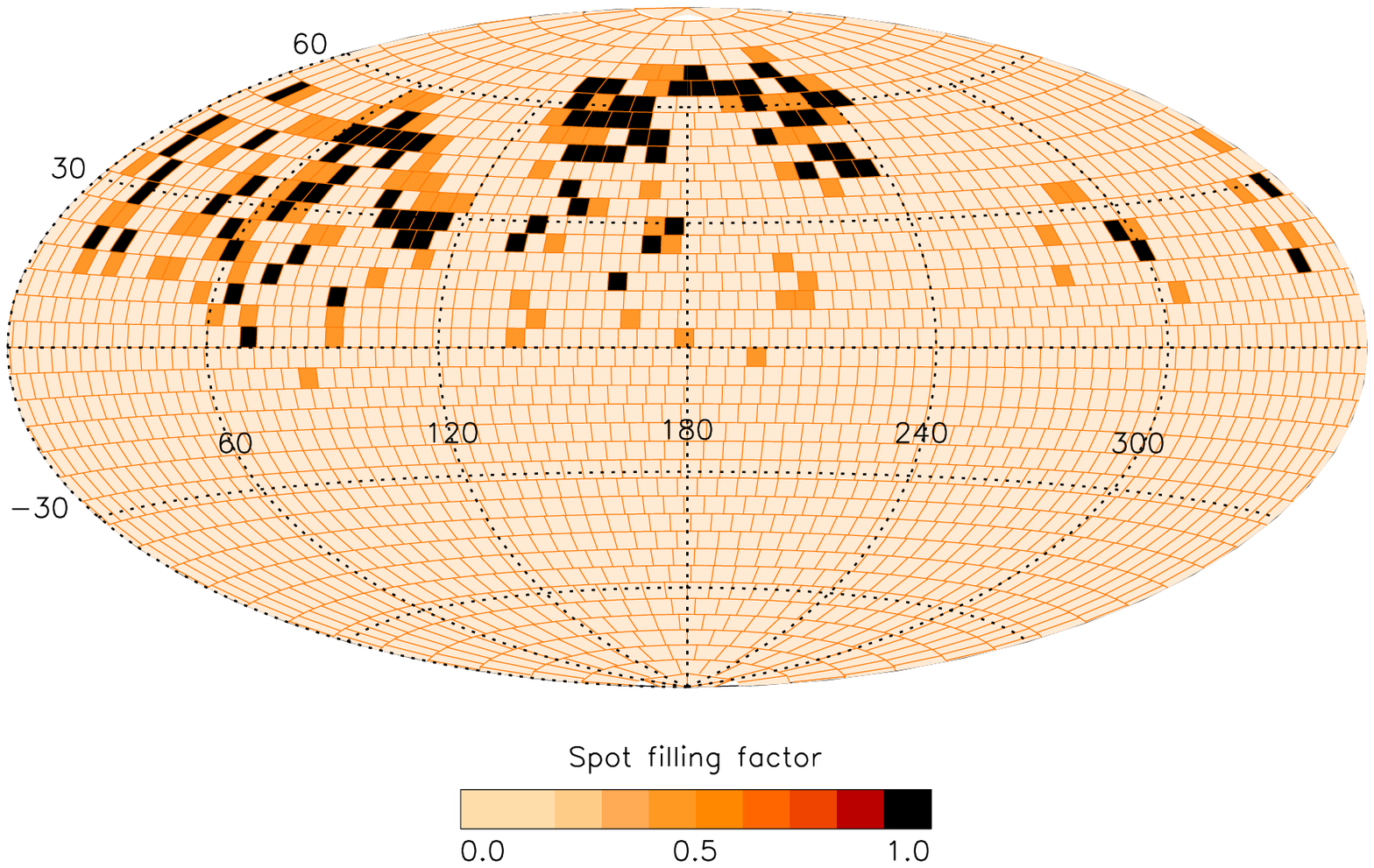}   
 } &
 \parbox{0.45\linewidth}{
  \center{\textsf{August 2 - ``6400~\AA''}} \\ 
  \vspace*{-0.5cm}
  \includegraphics*[width=\linewidth,clip=]{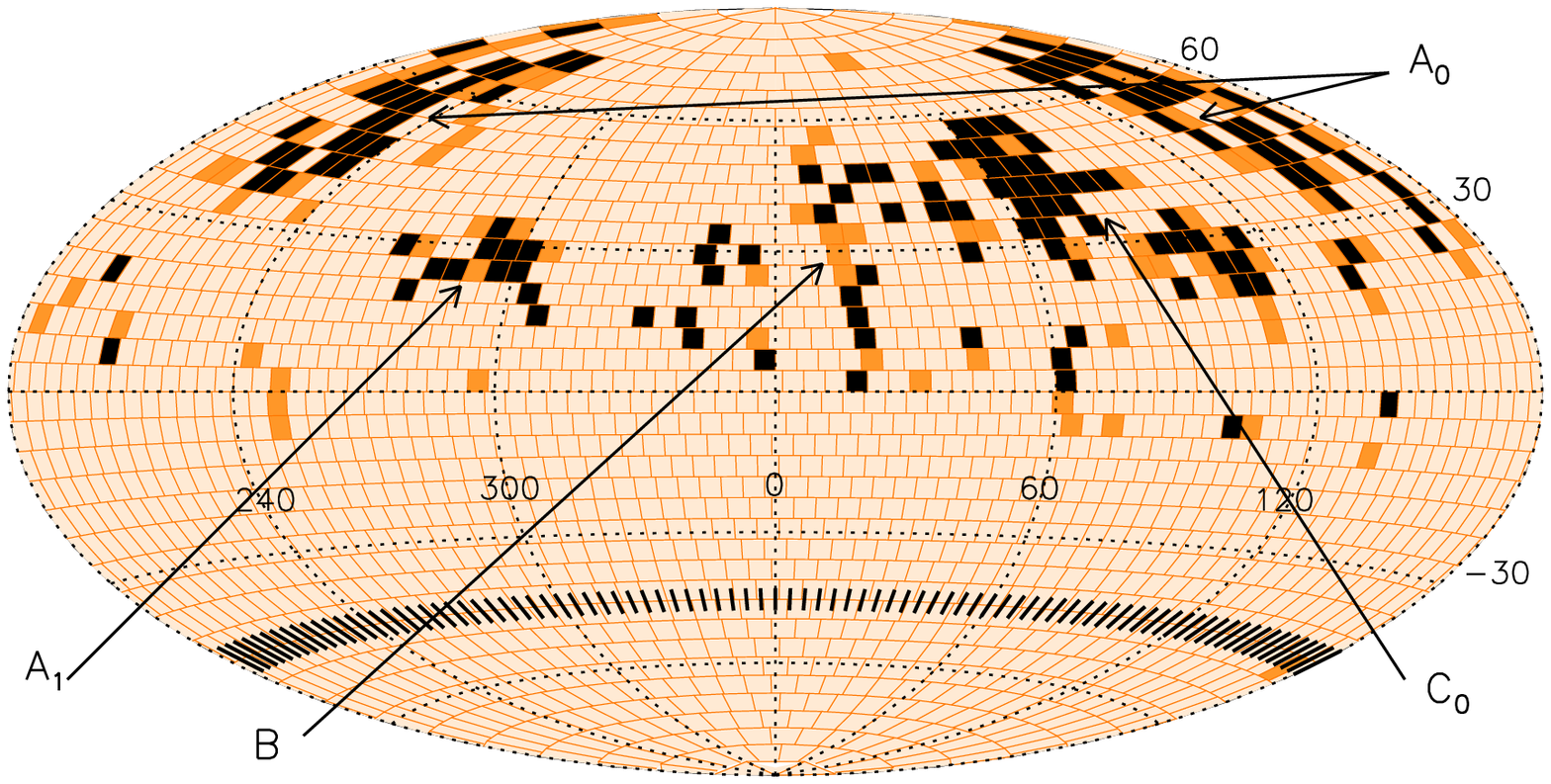}
  \includegraphics*[width=\linewidth,clip=]{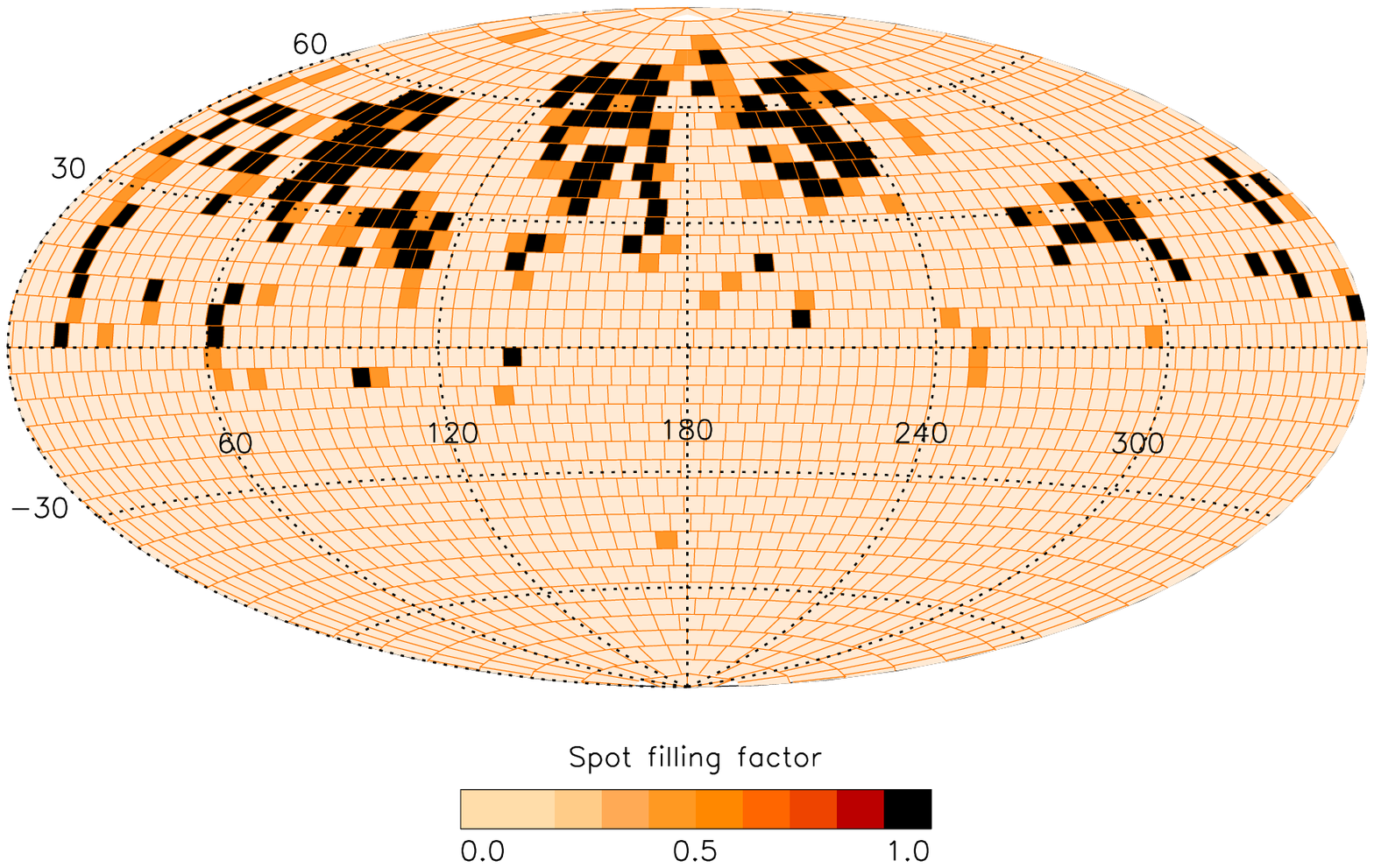}
 } \\

 \noalign{\medskip}
 \hline
 \noalign{\medskip}

\vspace{0.2cm}
 \parbox{0.45\linewidth}{
  \center{\textsf{August 7 - ``6120~\AA'' }} \\ 
  \vspace*{-0.5cm}
  \includegraphics*[width=\linewidth,clip=]{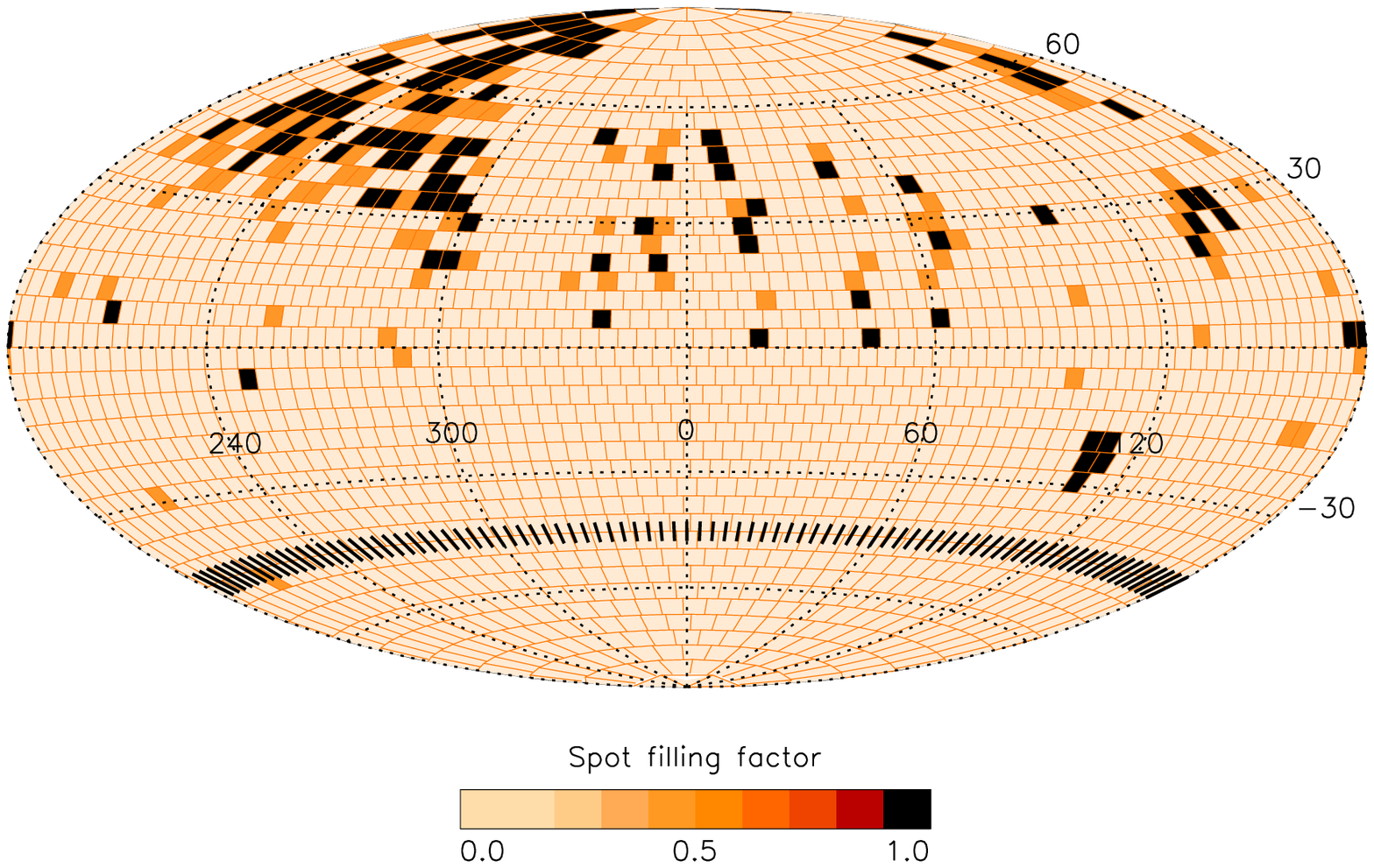}
  \includegraphics*[width=\linewidth,clip=]{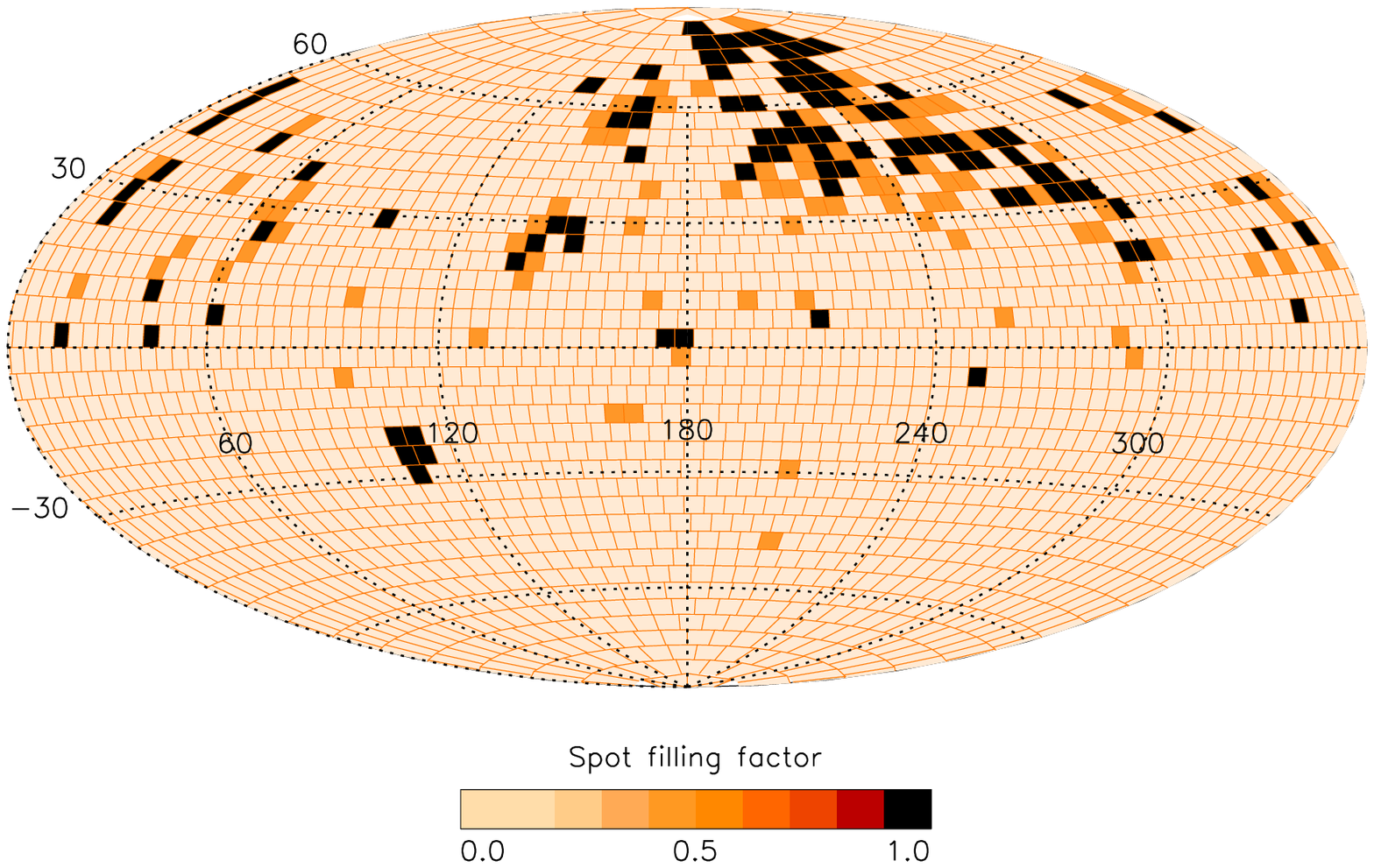}
  } &
 \parbox{0.45\linewidth}{
  \center{\textsf{August 7 - ``6400~\AA''}} \\ 
  \vspace*{-0.5cm}
  \includegraphics*[width=\linewidth,clip=]{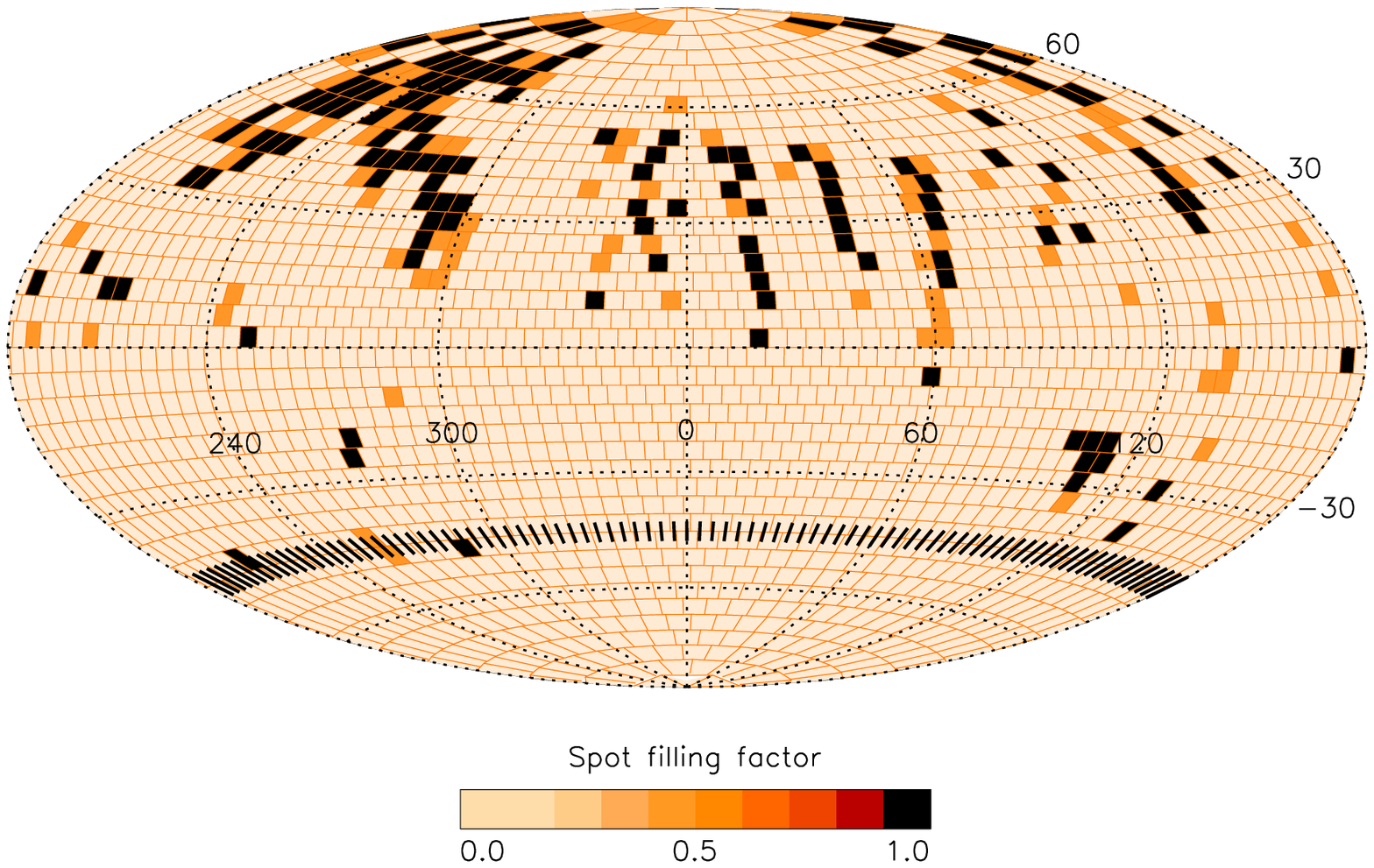}
  \includegraphics*[width=\linewidth,clip=]{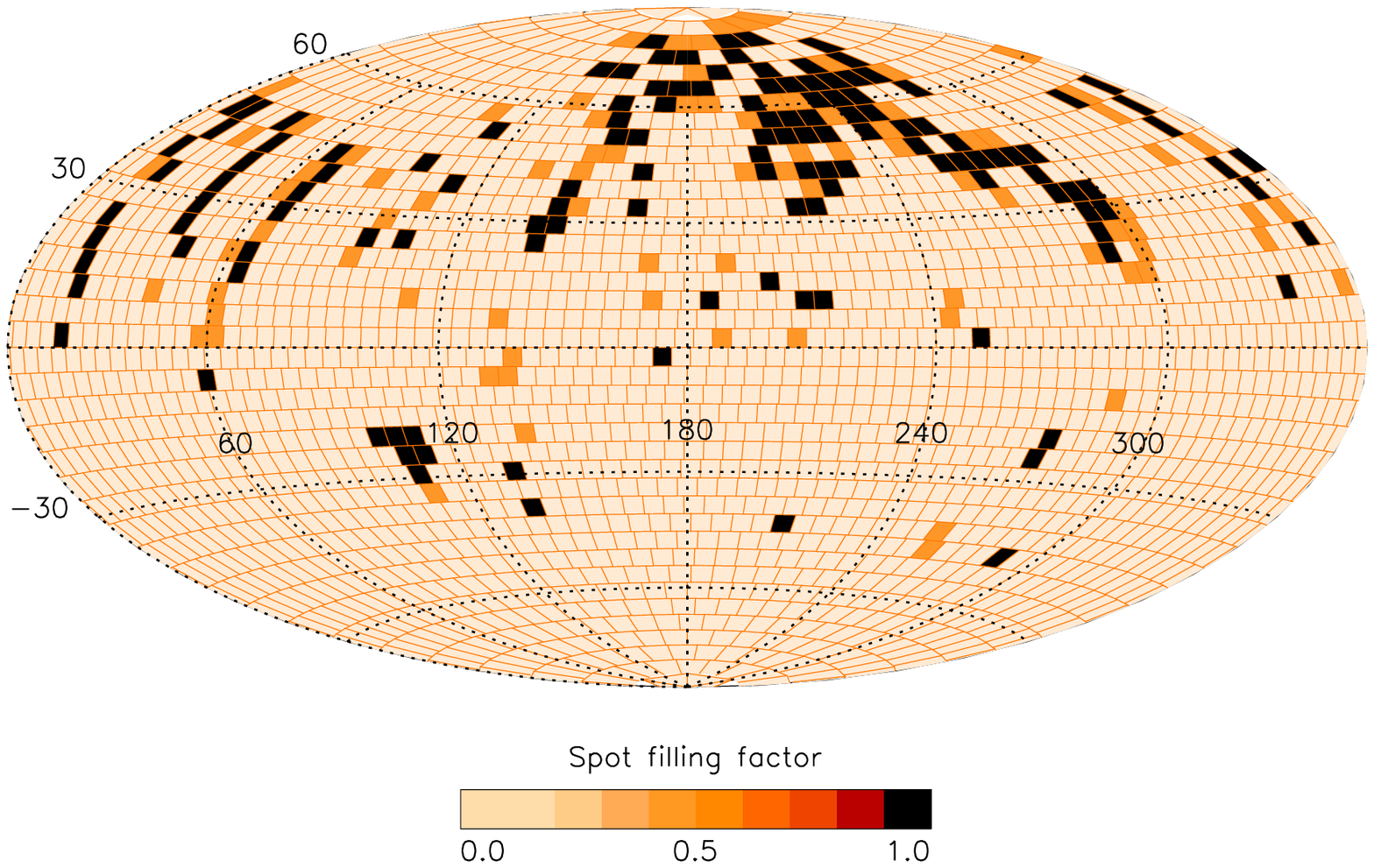}
 } 
\end{tabular}
 \medskip
 \caption{
          Doppler images of BO~Mic, reconstructed by CLDI 
          from the wavelength region given above each panel.
          The lower and upper map of each panel show the same surface, but rotated by 180~degrees.
          Black and dark~gray/orange areas represent surface regions completely covered or
          50\% covered with spots, respectively. Unspotted regions are rendered as light shades. 
          The subobserver longitudes of the observed phases are marked by short lines.
          A rotation phase $\phi=0$ corresponds to a subobserver longitude of 
          \mbox{$\varphi=0\degr$};
          rotation proceeds with \textit{decreasing} subobserver longitude.
          \mbox{About 13~stellar} rotations have passed between the images of the upper and the lower panels.
         }
 \label{fig:SM-CLDI-RMxTSA_HA-P0380-truecoords-1st2nd}
\end{figure*}

\subsection{Surface resolution}
\label{sec:resol}
The resolution achieved by a Doppler image depends on the phase sampling, the spectral resolution 
and the noise of the spectra used for its reconstruction.
Due to the one-dimensional projection of the surface structures onto each individual line profile,
the resolution also depends on the reconstructed latitude.
Although analytic estimates exist \cite[e.g.][]{Jankov92a} they are not much in use, because
the quantitative influence of the noise on the image resolution depends 
e.g. on the adopted atmospheric parameters and
requires simulations based on synthetic input data.
In addition, potential errors of the line profile modelling, imperfectly treated line blends
etc. can severely deteriorate the reconstruction quality. 
An extensive series of such simulations can be found in \citet{Rice01}.
We have studied the behaviour of CLDI based on synthetic input data
(\citealt{Wolter04}; 2005, in preparation).
It has already been extensively studied for CLDI's predecessor in \citet{Kuerster93}
with essentially the same results as for maximum entropy Doppler imaging. 

For the purpose of this paper, the resolution estimate is based on a comparison of the two
presented sets of images. They are reconstructed from two disconnected spectral regions of
rather different character (concerning the number of lines and the 
degree of line blending they contain). 
As a result, they can be considered as quite independent input data describing the same
surface patterns.

\begin{figure}
 \center{ 
  \hspace*{0.25cm} \\                  
  \includegraphics*[width=0.85\linewidth,clip=]{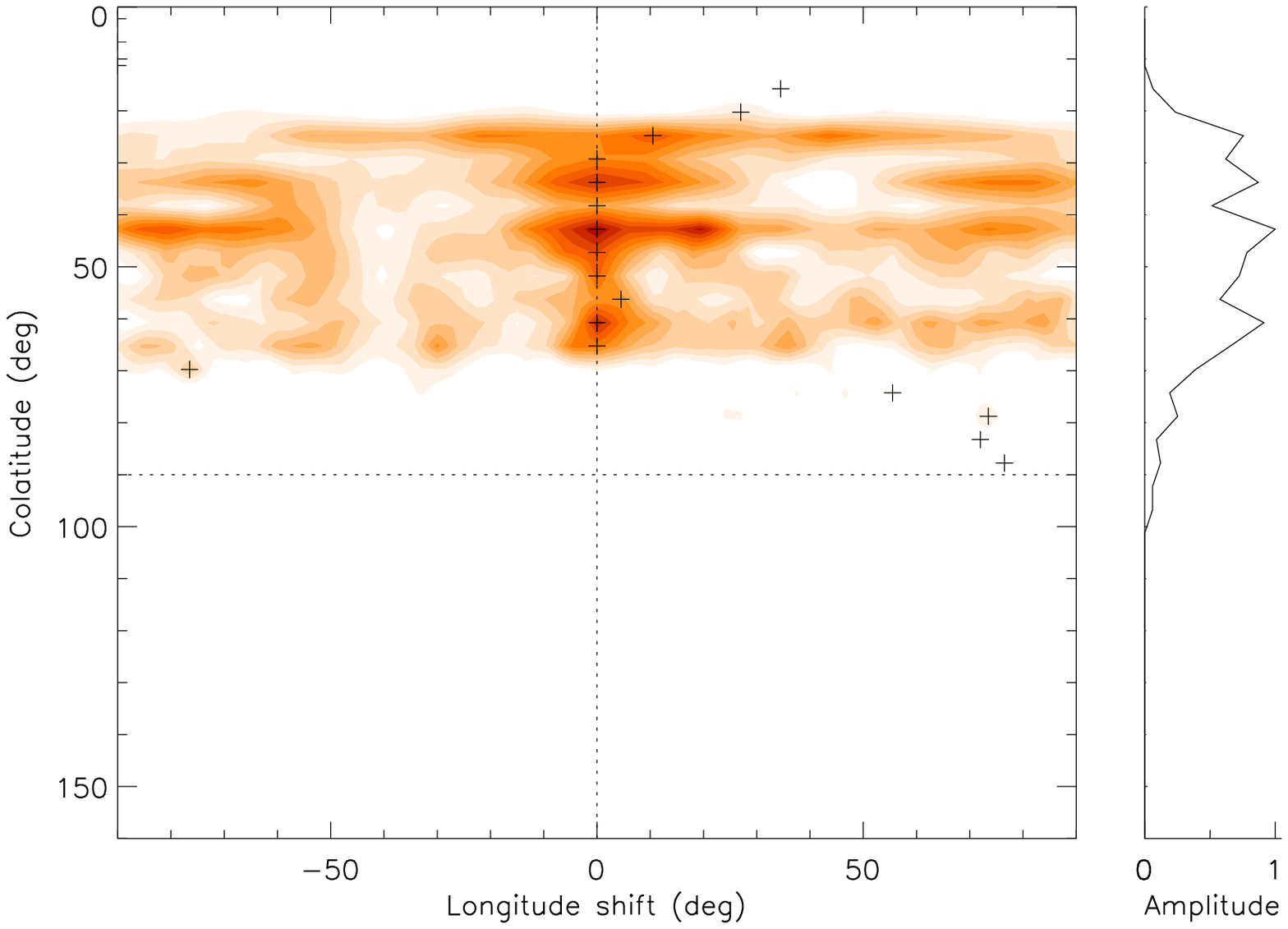}   
        }
   \caption{
   Quantifying the Doppler image resolution:
   Cross-correlation of different-wavelength Doppler images \textit{reconstructed from the
   same input spectra} (2004 August\,2, i.e. the 
   upper maps of Fig.~\ref{fig:SM-CLDI-RMxTSA_HA-P0380-truecoords-1st2nd}).
   The left panel shows the cross-correlation of corresponding latitude strips of the compared maps;
   darker shades represent larger values.
   The relative maximum on each colatitude \mbox{($= 90\degr -$ latitude)} is marked by a plus, the values of these maxima
   are plotted in the right panel. See text for details. 
           }
    \label{fig:SM_DI_resolution-xcorr}
\end{figure}

The images of Fig.~\ref{fig:SM-CLDI-RMxTSA_HA-P0380-truecoords-1st2nd} belonging 
to the same input spectra (i.e. to the same dates) show corresponding features on scales down to about
\mbox{10\degr$\times$10\degr} on the surface; the correspondence becomes poorer
within about 15\degr\ from the pole and the equator.
This somewhat subjective estimate is substantiated by the surface cross-correlation \citep[e.g.][]{Tonry79}
shown in Fig.~\ref{fig:SM_DI_resolution-xcorr}:
The correlation peaks are located within $\pm10\degr$ in the given latitude range. 
The cross-correlation
was computed along longitude for each latitude strip of the compared Doppler maps.

\section{Results}

\begin{figure}  
\center{
  \vspace*{0.2cm}
    \includegraphics*[width=\linewidth,clip=]{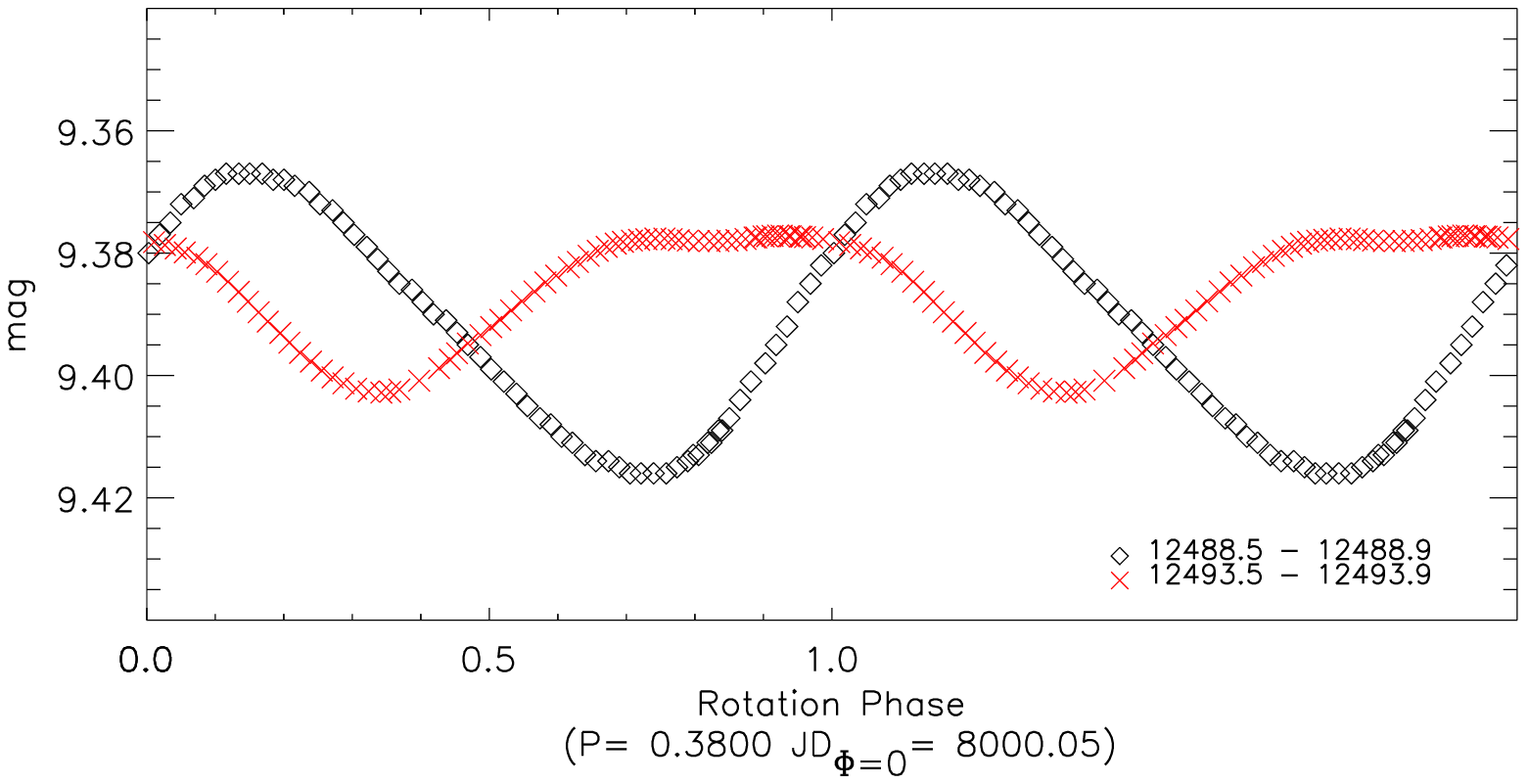}               
  \vspace*{-0.6cm}
       }
  \caption{
           Lightcurves computed from Doppler images of the ``6400~\AA'' wavelength region.
           The symbols represent the August~2~($\diamond$) and August~7~(\small{$\times$}) reconstructions,
           respectively.
           The lightcurves are computed from the corresponding surfaces shown in 
           Fig.~\ref{fig:SM-CLDI-RMxTSA_HA-P0380-truecoords-1st2nd}.
           The annotation dates are JD-2440000.   
          }
  \label{fig:DI-lcurves_6400-P0380} 

  \bigskip
\center{
    \includegraphics*[width=\linewidth,clip=]{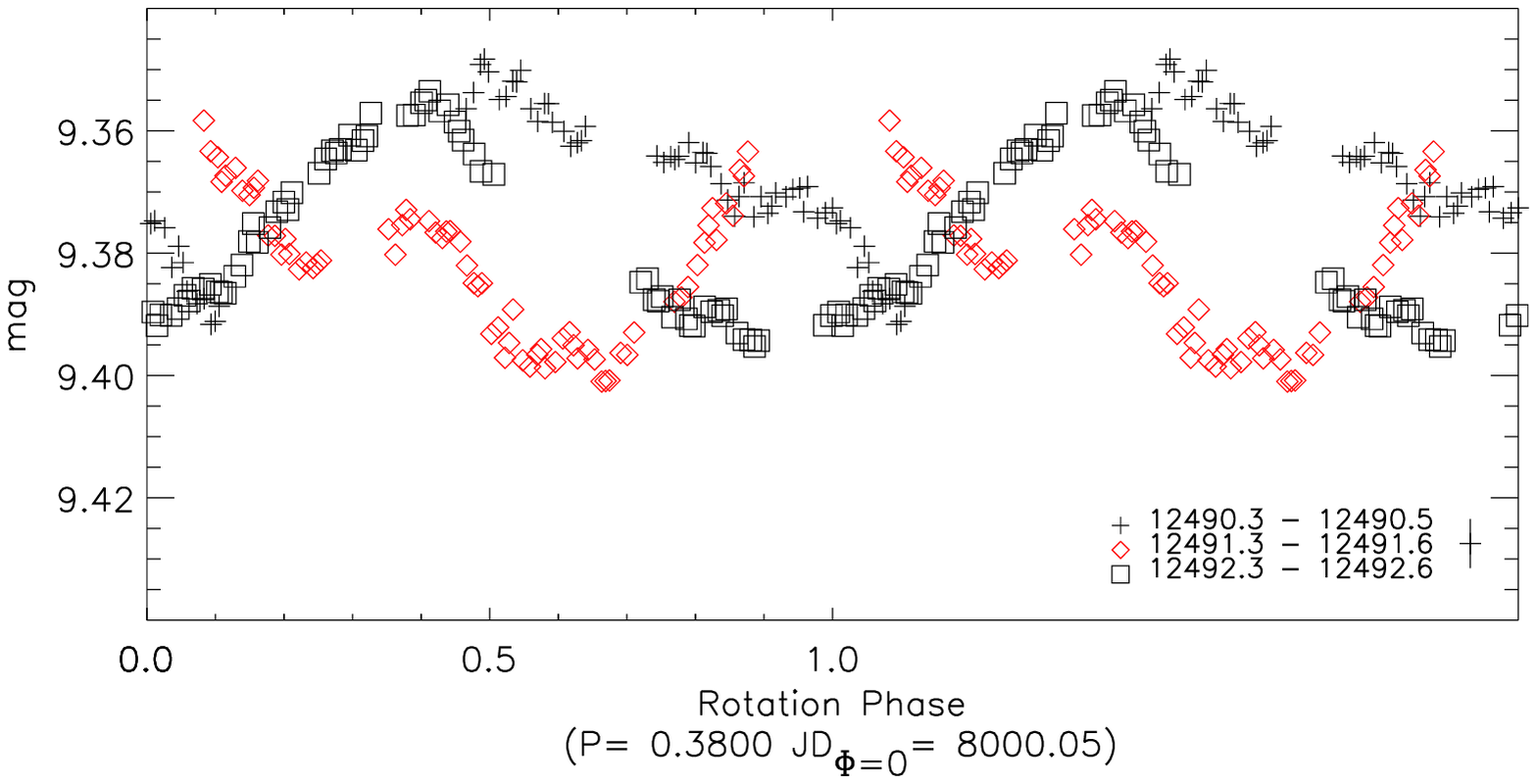}
  \vspace*{-0.6cm}
       }
  \caption{
    Lightcurve of BO~Mic (2002 August 3-5), folded with a period of 0.380~days.
    Note that this folding period does not merge any portion of the data observed
    during different nights into a unique lightcurve.
    Annotation dates are JD-2440000:
    $+$~=~August~3, $\diamond$~=~August~4 and {\tiny$\Box$}~=~August~5.
          }
  \label{fig:lcurves_folded-SAAOphot_P0380} 
\end{figure}

\subsection{Observed spot evolution}
Comparing the August~2 and August~7 images, 
several spot groups are recognizable  (Fig.~\ref{fig:SM-CLDI-RMxTSA_HA-P0380-truecoords-1st2nd}):
\begin{list}{$\bullet$}{\setlength {\itemsep}{2pt}
                      \setlength {\topsep}{2pt} }  
  \item[$\mathsf{A_0}$:]{
    A large irregularly shaped spotted area extending from about 
    120\degr\ to 240\degr\ in longitude and from
    about 30\degr\ up to above 60\degr\ in latitude
            }
  \item[$\mathsf{A_1}$:]{
    A smaller spot (group) centered slightly below 30\degr\ latitude and at about
    290\degr\ longitude   
       }
  \item[$\mathsf{B}$:]{
    An extended area sparsely covered with spots, centered close to 0\degr\
    longitude and 30\degr\ latitude; its shape is reminiscent of a ``$\lambda$''
             }
  \item[$\mathsf{C_0}$:]{
    A spot group extending from about 50\degr\ to 120\degr\ in longitude and
    from close to the equator to about 60\degr\ in latitude.
    In the August~2 images this group is far more pronounced, showing
    a separation in two distinct, apparently quite coherent groups.
             }
  \item[$\mathsf{C_s}$:]{
    The small coherent southern spot only visible in the  August~7 images
    at about 110\degr\ longitude and -25\degr~latitude. 
    It may be mirrored to the northern hemisphere in the August~2 images,
    appearing there as part of group~$\mathsf{C_0}$.
             }
\end{list}
\medskip

\noindent 
The following evolution of the spot pattern
appears to have taken place during the about~13 \mbox{($\frac{5~\mathrm{days}}{0.38~\mathrm{days}0})$} 
rotations between the August~2 and August~7 images:
Group $\mathsf{A_0}$ undergoes rather weak changes but is shifted in longitude by roughly
30\degr.
Equivalently, $\mathsf{A_0}$ could be said to stay in place and fade at lower 
longitudes while growing at its high longitude side.
Group $\mathsf{A_1}$ appears to stay in place, getting
connected to $\mathsf{A_0}$.
Group $\mathsf{B}$ remains largely unchanged, preserving its $\lambda$-shape
and growing a weak ``annex'' to
the north west at about 320\degr\ longitude  and 45\degr\ latitude. 
Finally, most of group $\mathsf{C_0}$ disappears with possibly one
component reappearing in the spot $\mathsf{C_s}$.

The apparent shift of group $\mathsf{A_0}$ towards increasing longitudes
(i.e. earlier rotation phases, keeping in mind that 
rotation takes place with \textit{decreasing} sub-observer longitude), 
combined with the pronounced weakening
of group  $\mathsf{C_0}$ at lower longitudes (i.e. later rotation phases)
leads to a ``darkening'' at earlier phases and a ``brightening'' at later ones.
As a result, the lightcurve minimum  is shifted to earlier phases when
going from August~2 to August~7; this 
can be seen in the synthetic lightcurves computed from the Doppler images,
shown in Figure~\ref{fig:DI-lcurves_6400-P0380}.

The available photometric lightcurve (2004 \mbox{August~3-5}) could not be observed
strictly parallel to the spectroscopic observations used for the Doppler
imaging;  
even though it was observed on three consecutive nights, 
it contains unobserved gaps each lasting about 1.5 rotations due to the short
rotation period of BO~Mic.
However, the available lightcurve
indicates that \textit{spot reconfigurations comparable to 
those seen in the Doppler images have occurred
on time scales as short as about two  rotations}
(namely between two consecutive nights). 
Due to the merely disk-integrated information 
about the spot pattern contained in the lightcurve, 
these spot reconfigurations cannot be further specified.

The inferred fast changes of some of BO~Mic's spots during the August~2002 observations
on the one hand, and the
snapshot-like information of the Doppler images (as well as the lightcurves)
on the other, make definite statements about the spot lifetimes difficult.
However, some inferences can be made:
\begin{list}{$\bullet$}{\setlength {\itemsep}{2pt}
                      \setlength {\topsep}{2pt} }  
  \item[(a)]{
     Several spot groups of irregular shape, typically extending 10-20\degr\ on the surface,
     located from equatorial up to about 60\degr\ latitudes, reappear quite
     well-preserved after 13 rotations.  
          }
  \item[(b)]{
     One quite coherent spot group of about 30\degr$\times$30\degr\ surface extension
     has substantially weakened after 13 rotations.
          }
  \item[(c)]{
     The only large spot group, quite coherent but of irregular shape, extends more 
     than 100\degr\ in longitude and about 50\degr\ in latitude. 
     It reappears after 13 rotations with a comparable surface extension but significantly
     shifted and transformed on scales of up to 30\degr.
          }
\end{list}
\medskip

\noindent A detailed comparison between our Doppler images of BO Mic and those of
\citet{Barnes01} is difficult. The reliability of small-scale features
 in their images is presumably low, because of partly low SNR and
 partly rather inhomogeneous phase coverage of their input spectra.
 However, the large-scale distribution of spots does show
 similarities. Barnes et al.'s 1998 images of BO Mic, as well as our
 2002 images presented here, both show spots over a wide range of
 latitudes including regions close to the equator.
Both sets of images show one large spot group extending
 up to or very close to the poles, but neither the 1998 images nor the
 2002 images exhibit a polar spot.
Our 2002 images clearly show that the hemisphere opposite to the
mentioned large spot group is considerably less densely spotted;
this is supported by the modulation of the observed lightcurves  
(Fig.~\ref{fig:lcurves_folded-SAAOphot_P0380}). 
Two differently spotted hemispheres also exist in
some of Barnes et al.'s 1998 images. However, we think that due to the partly
unfavourable phase sampling of the 1998 observations of BO~Mic, no
definite statement can be made about this.

It must be kept in mind that all identified spot groups found in the 
Doppler images may comprise groups truly located on both stellar hemispheres, 
incorrectly reconstructed only on the northern hemisphere by the Doppler 
imaging.

\subsection{Differential rotation?}

\begin{figure}
 \center{ 
   \includegraphics*[width=0.85\linewidth,clip=]{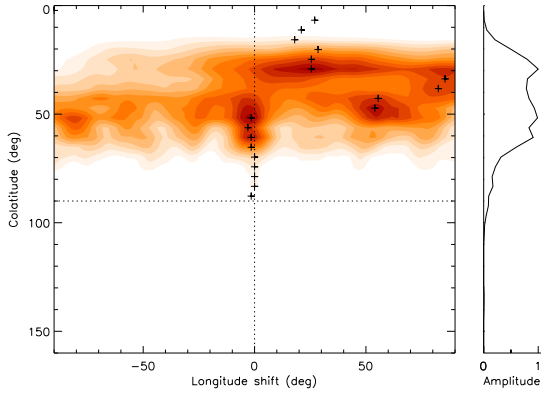}   
        }
   \caption{
   Cross-correlation of the August~2 and August~7 Doppler images of BO~Mic,
   reconstructed from the ``6400~\AA'' wavelength region
   (right column maps of Fig.~\ref{fig:SM-CLDI-RMxTSA_HA-P0380-truecoords-1st2nd}).
   The compared images have been smoothed by a two-dimensional boxcar extending 
   \mbox{10\degr$\times$10\degr} to account for their surface resolution
   estimated  above.  
           }
    \label{fig:SM_DI-1st_xcorr_2nd}
\end{figure}

The intermediate to high latitude spot 
group~$\mathsf{A_0}$ is apparently shifted to larger longitudes (i.e. against the direction of rotation)
by about 30\degr,
while features closer to the equator (including group $\mathsf{A_1}$) 
roughly remain stationary;
this result of a visual inspection is confirmed by the surface cross-correlation shown in
Fig.~\ref{fig:SM_DI-1st_xcorr_2nd}
(see Sec.~\ref{sec:resol} for an explanation of the cross-correlation procedure).
However, the surface shear visible in the cross-correlation map is not at all 
smoothly increasing from the equator to the poles as would be the case for a differential 
rotation law comparable to the solar case (Eq.~\ref{eq:sunspot_rotation_law}).
Instead, following the discussion of the preceeding section, 
we suggest that the apparent surface shear 
is predominantly due to
the intrinsic evolution of the individual spot groups, i.e. their appearance and decay.

As a result our 2004 Doppler images of BO~Mic only allow us to
estimate an upper limit of the
differential rotation strength which could be veiled by the intrinsic spot evolution:
Inspecting Fig.~\ref{fig:SM_DI-1st_xcorr_2nd} yields a pole-equator shear estimate
of  \mbox{$\Delta\varphi\approx -20\pm10\degr\ $}.
The estimated scatter of $\pm10\degr\ $ corresponds well with the resolution estimate of
our Doppler images made in Sec.~\ref{sec:resol}.
This translates into a ``relative strength'' (cf. Eq.~\ref{eq:diffrot-alpha_definition}) 
of the differential rotation of 
\begin{equation}
  \vert \alpha \vert <  \Big\vert \frac{-20\pm10\deg}{13.2\cdot360\deg} \Big\vert 
                      =  0.004 \pm 0.002  \quad
\label{eq:SM_iii_diffrot_upplim-estim}
\end{equation}
i.e. at most 50~times smaller than the solar value.


\section{Summary and Conclusions}
\begin{figure}  
   \epsfig{file=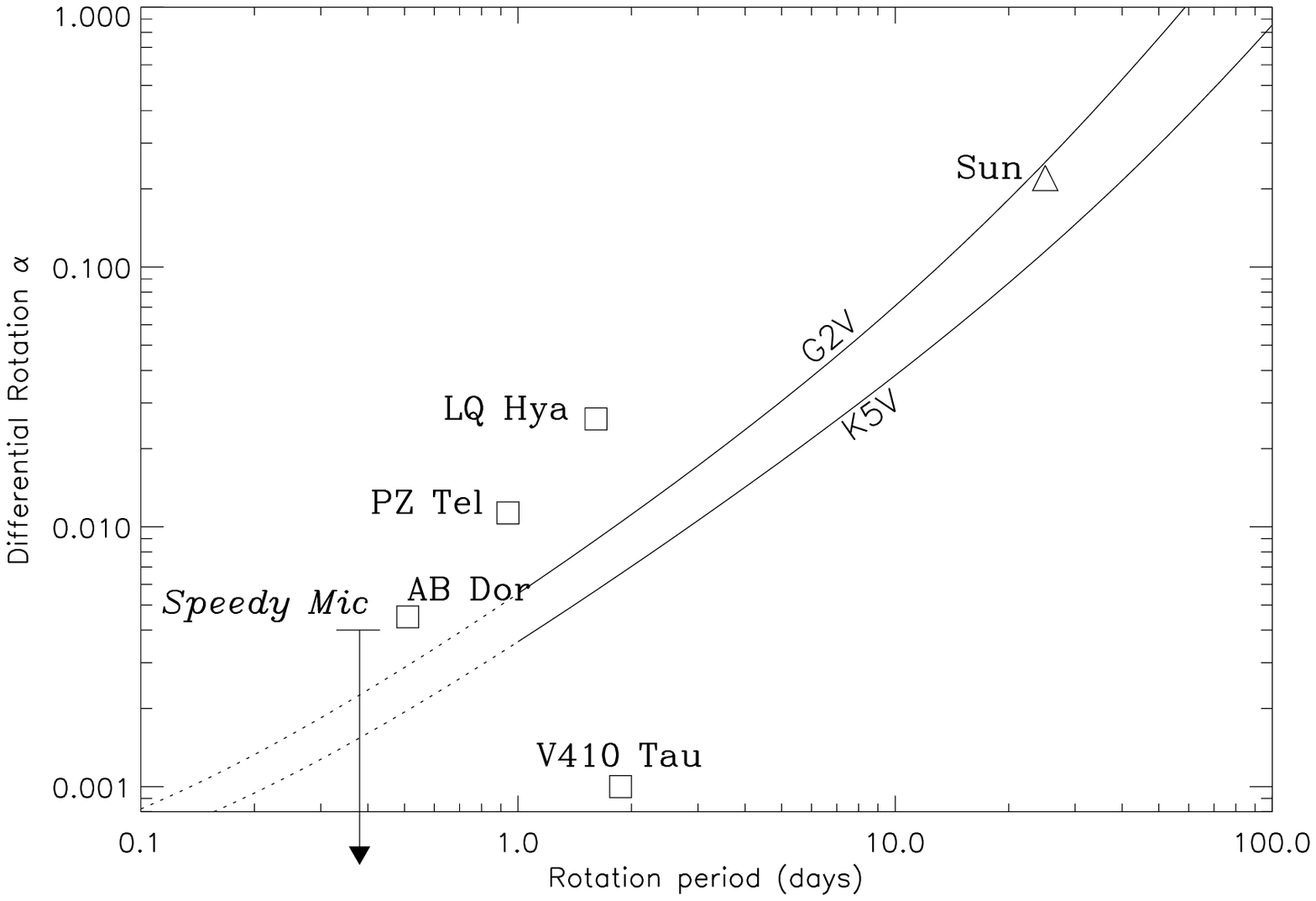,width=0.99\linewidth,clip=}
  \caption{
  Theory and measurements of differential rotation of (apparently) single dwarf stars.
  The curves are adapted from \citet{Kitchatinov99} and extrapolated below a period of one~day.
  Squares mark measurements 
  based on comparisons of Doppler images 
  \citep{Rice96,Barnes00,Donati99,Donati03}
  or the related 
  method of ``direct spot tracking'' \citep{CollierCameron02}.
  The upper limit 
  determined for BO~Mic (Speedy Mic) in this 
  work is rendered by
  the arrow's top. The arrowhead drawn outside the plot 
  illustrates the possibility of \textit{anti-solar} differential rotation 
  (i.e. the equator rotating slower than the poles in terms of $\Omega$)
  which is not excluded by our observations.
          } 
  \label{fig:Obs_difffrot_non-solar-andmodel}
\end{figure}

We have acquired a time series of high-resolution spectra of BO~Mic
which is of outstanding quality regarding its low noise level and
its dense and even sampling of rotation phases.
BO~Mic shows an exceptionally short rotation period which makes it
a very interesting object for studying the spot evolution and possibly the
differential rotation of ultrafast rotators (UFR). 

Our spectra allow the reconstruction of two Doppler images of BO~Mic,
seperated by about
13~stellar rotations.
Each of these images is based on spectra observed during a single stellar rotation.
This minimizes the influence of intrinsic spot evolution on the images,
which 
must be considered a possibility
even during a single rotation. 

We have applied our spectrum deconvolution algorithm sLSD to extract line profiles
from two separate wavelength ranges of the spectra.
The Doppler images reconstructed from the resulting two 
independent input datasets
agree on scales down to \mbox{10\degr$\times$10\degr}\ on the stellar 
surface, excluding latitudes near the pole and the equator.
In this way we estimate the surface resolution of our images to about~$10\degr$.

While the large-scale spot pattern has survived the 13~rotations covered by our
images, several reconfigurations of spots have definitely 
taken place during that time span.
These reconfigurations extend up to about \mbox{30\degr$\times$30\degr}\
on the surface.

While our Doppler images show the timescale of
these spot reconfigurations to be shorter than 13~rotations of BO~Mic,
our nearly simultaneous photometric observations suggest that similar 
reconfigurations have even taken place between consecutive observing nights,
i.e. during about 2.5~stellar rotations. 

Currently, for stars other than the Sun
only very few
observations of spot lifetimes 
on timescales of a few stellar rotations exist.
By far the best-studied object concerning short-term spot evolution
is the Sun.
Compared to the solar case, ``fast'' changes of the spot pattern are not surprising:
Pratically no sunspot group survives more than two solar rotations and definitely not
without major reconfigurations.

The observed spectral line profiles of BO~Mic contain 
remarkably rich information
which agrees between the different wavelength ranges. 
This fact and the resulting
quality of the Doppler images 
make our observations highly successful.
However, one of our major aims could not be achieved, namely the 
observation of differential rotation on BO~Mic
and the determination of its strength.
Given the large-scale stabilty of the spot pattern deduced from our Doppler images,
the differential rotation can safely be said to be weak
compared to the Sun.
Unfortunately, further determination of the differential rotation strength is
massively impaired by the above-discussed intrinsic evolution of the observed spots.

Our determined upper limit of \mbox{$\vert\alpha\vert < 0.004\pm0.002$} for the strength
of the differential rotation agrees well with a previously determined value by
\citet{Barnes01}.
Combined with measurements of differential rotation on other fast rotating dwarf stars,
shown in Fig.~\ref{fig:Obs_difffrot_non-solar-andmodel},
it confirms the observed pronounced decline
of differential rotation strength with decreasing rotation periods.  
As Fig.~\ref{fig:Obs_difffrot_non-solar-andmodel} also illustrates, this is
well in accord with the rotational models of \citet{Kitchatinov99},
so far supporting the mean-field-based modelling approach for the
convection zones of solar-like stars.

\begin{acknowledgements}
      U.W. acknowledges financial support from
      \emph{Deut\-sche For\-schungs\-ge\-mein\-schaft}, \mbox{DFG - SCHM 1032/9-1}.
      We thank D. Kilkenny for making the photometric observations at SAAO possible
      and G.~Cutispoto for kindly supplying photometric data of Speedy Mic.
      Finally, we thank our referee J. Barnes for his careful reading of the manuscript
      and his concise suggestions which helped to improve it. 
\end{acknowledgements}

\end{document}